%% file: paper3.tex
\documentclass[longauth]{aa}  
\usepackage{graphicx}
\usepackage{txfonts}
\usepackage{natbib}
\bibpunct{(}{)}{;}{a}{}{,} % to follow the A&A style

\usepackage[colorlinks=true,allcolors=blue]{hyperref}
\usepackage{multirow}
\usepackage{multicol}
\usepackage{placeins}

\newcommand{\ms}{$M_{\odot}$}
\newcommand{\rs}{$R_{\odot}$}
\newcommand{\mzams}{$M_{\rm ZAMS}$}
\newcommand{\mej}{$M_{\rm ej}$}
\newcommand{\mhenv}{$M_{\rm H,env}$}
\newcommand{\mh}{$M_{\rm H}$}
\newcommand{\ra}{$R$}
\newcommand{\e}{\emph{E}}
\newcommand{\Ni}{{$^{56}$Ni}}
\newcommand{\mni}{$M_{\rm Ni}$}
\newcommand{\mix}{$^{56}\rm Ni$ mixing}
\newcommand{\texp}{$t_{\rm exp}$}
\newcommand{\sneii}{SNe~II}
\newcommand{\snii}{SN~II}
\newcommand{\cd}{Cd}
\newcommand{\pd}{pd}
\newcommand{\optd}{optd}
\newcommand{\mbolend}{$M_{\rm bol,end}$}
\newcommand{\mboltail}{$M_{\rm bol,tail}$}
\newcommand{\sone}{$s_{\rm 1}$}
\newcommand{\stwo}{$s_{\rm 2}$}
\newcommand{\sthr}{$s_{\rm 3}$}
\newcommand{\aeha}{$a/e$}
\newcommand{\ha}{H$_{\rm \alpha}$}
\newcommand{\hb}{H$_{\rm \beta}$}

\newcommand{\gr}{$(g-r)$}
\newcommand{\tpt}{$t_{\rm PT}$}
\newcommand{\ttrans}{$t_{\rm trans}$}
\newcommand{\vphfi}{$v_{\rm ph,50}$}
\newcommand{\p}{$\rho$}

\defcitealias{anderson+14_lc}{A14}
\defcitealias{gutierrez+17II}{G17}
\defcitealias{dejaeger+18}{dJ18}
\defcitealias{martinez+20}{M20}
\defcitealias{martinez+21a}{Paper~I}
\defcitealias{martinez+21b}{Paper~II}

\usepackage{color}

\begin{document}

        \title{Type II supernovae from the Carnegie Supernova Project-I}
        \subtitle{III. Understanding SN~II diversity through correlations between physical and observed properties}
        
                \author{L. Martinez \inst{1,2,3}
                \and J.~P. Anderson \inst{4}
        \and M.~C. Bersten \inst{1,2,5}
        \and M. Hamuy \inst{6,7}
        \and S. González-Gaitán \inst{8}
        \and M. Orellana \inst{3,9}
        \and \\ M. Stritzinger \inst{10}
        \and M.~M. Phillips \inst{11}
        \and C.~P. Guti\'errez \inst{12,13}
        \and C.~Burns \inst{14}
        \and T. de Jaeger \inst{15,16}
        \and K. Ertini \inst{1,2}
        \and G. Folatelli \inst{1,2,5}
        \and \\ F. Förster \inst{17,18,19,20}
        \and L. Galbany \inst{21}
        \and P. Hoeflich \inst{22}
        \and E.~Y. Hsiao \inst{22}
        \and N. Morrell \inst{11}
        \and P.~J. Pessi \inst{2,4}
        \and \\ N.~B. Suntzeff \inst{23}
        }

        \institute{Instituto de Astrof\'isica de La Plata (IALP), CCT-CONICET-UNLP. Paseo del Bosque S/N, B1900FWA, La Plata, Argentina \\
                        \email{laureano@carina.fcaglp.unlp.edu.ar}
                \and
                        Facultad de Ciencias Astron\'omicas y Geof\'isicas, Universidad Nacional de La Plata, Paseo del Bosque S/N, B1900FWA, La Plata, Argentina
                \and
            Universidad Nacional de R\'io Negro. Sede Andina, Mitre 630 (8400) Bariloche, Argentina
        \and
            European Southern Observatory, Alonso de Córdova 3107, Casilla 19, Santiago, Chile
        \and
            Kavli Institute for the Physics and Mathematics of the Universe (WPI), The University of Tokyo, 5-1-5 Kashiwanoha, Kashiwa, Chiba 277-8583, Japan
        \and
            Vice President and Head of Mission of AURA-O in Chile, Avda. Presidente Riesco 5335 Suite 507, Santiago, Chile
        \and
            Hagler Institute for Advanced Studies, Texas A\&M University, College Station, TX 77843, USA
        \and
            CENTRA-Centro de Astrofísica e Gravitaçäo and Departamento de Física, Instituto Superio Técnico, Universidade de Lisboa, Avenida Rovisco Pais, 1049-001 Lisboa, Portugal
        \and
            Consejo Nacional de Investigaciones Cient\'ificas y T\'ecnicas (CONICET), Argentina.
        \and
            Department of Physics and Astronomy, Aarhus University, Ny Munkegade 120, DK-8000 Aarhus C, Denmark
        \and
            Carnegie Observatories, Las Campanas Observatory, Casilla 601, La Serena, Chile
        \and 
            Finnish Centre for Astronomy with ESO (FINCA), FI-20014 University of Turku, Finland
        \and
            Tuorla Observatory, Department of Physics and Astronomy, FI-20014 University of Turku, Finland
        \and 
            Observatories of the Carnegie Institution for Science, 813 Santa Barbara St., Pasadena, CA 91101, USA
        \and
            Institute for Astronomy, University of Hawaii, 2680 Woodlawn Drive, Honolulu, HI 96822, USA
        \and
            Department of Astronomy, University of California, 501 Campbell Hall, Berkeley, CA 94720-3411, USA
        \and
            Data and Artificial Intelligence Initiative, Faculty of Physical and Mathematical Sciences, University of Chile, Santiago, Chile
        \and
            Centre for Mathematical Modelling, Faculty of Physical and Mathematical Sciences, University of Chile, Santiago, Chile
        \and
            Millennium Institute of Astrophysics, Santiago, Chile.
        \and
            Department of Astronomy, Faculty of Physical and Mathematical Sciences, University of Chile, Santiago, Chile
        \and
            Institute of Space Sciences (ICE, CSIC), Campus UAB, Carrer de Can Magrans, s/n, E-08193 Barcelona, Spain
        \and
            Department of Physics, Florida State University, 77 Chieftan Way, Tallahassee, FL 32306, USA
        \and
            George P. and Cynthia Woods Mitchell Institute for Fundamental Physics and Astronomy, Department of Physics and Astronomy, Texas A\&M University, College Station, TX 77843
            }

\titlerunning{SN~II diversity through correlations between physical and observed properties}

%\authorrunning{}

\date{Received XXX; accepted XXX}
 
%===================================================
% ABSTRACT
%===================================================

\abstract 
{Type II supernovae (\sneii) show great photometric and spectroscopic diversity which is attributed to the varied physical characteristics of their progenitor and explosion properties.
In this study, the third of a series of papers where we analyse a large sample of \sneii\ observed by the Carnegie Supernova Project-I, we present correlations between their observed and physical properties.
Our analysis shows that explosion energy is the physical property that correlates with the highest number of parameters.
We recover previously suggested relationships between the hydrogen-rich envelope mass and the plateau duration, and find that more luminous \sneii\ with higher expansion velocities, faster declining light curves, and higher \Ni\ masses are consistent with higher energy explosions. In addition, faster declining \sneii\ (usually called SNe~IIL) are also compatible with more concentrated \Ni\ in the inner regions of the ejecta.
Positive trends are found between the initial mass, explosion energy, and \Ni\ mass. While the explosion energy spans the full range explored with our models, the initial mass generally arises from a relatively narrow range.
Observable properties were measured from our grid of bolometric LC and photospheric velocity models to determine the effect of each physical parameter on the observed \snii\ diversity.
We argue that explosion energy is the physical parameter causing the greatest impact on \snii\ diversity, that is, assuming the non-rotating solar-metallicity single-star evolution as in the models used in this study.
The inclusion of pre-SN models assuming higher mass loss produces a significant increase in the strength of some correlations, particularly those between the progenitor hydrogen-rich envelope mass and the plateau and optically thick phase durations.
These differences clearly show the impact of having different treatments of stellar evolution, implying that changes in the assumption of standard single-star evolution are necessary for a complete understanding of \snii\ diversity.}


\maketitle
%------------------------------------------------------
\section{Introduction} 
\label{sec:intro} 

Type II supernovae (\sneii\footnote{Throughout this paper we use `\sneii' to refer to all hydrogen-rich core-collapse SNe that show slow- and fast-declining light curves (historically referred to as SNe~IIP and IIL, respectively), excluding type~IIb, IIn, and SN~1987A-like events.}) result from the terminal explosion of massive stars (>8$-$10~\ms) that have retained a significant fraction of their hydrogen-rich envelope at the time of core collapse.
\sneii\ are classified by the presence of prominent hydrogen lines in their spectra \citep{minkowski41} and are the most common type of core-collapse SN in nature \citep{li+11a,shivvers+17}.
The direct detection of progenitors of \sneii\ in pre-explosion images provides increasing evidence for red supergiant (RSG) progenitors \citep[e.g.][]{vandyk+03,smartt09} with initial masses in the range of $\sim$8$-$18~\ms\ \citep{smartt15}.

Statistical studies of \sneii\  have revealed the great diversity in their photometric and spectral properties.
In addition, a continuum of light-curve (LC) parameters (e.g. absolute magnitudes, declination rates, time durations of different phases), colours, expansion velocities, and equivalent widths of a number of spectral lines are observed \citep[e.g.][]{hamuy03,bersten13phd,anderson+14_lc,sanders+15,valenti+16,gutierrez+17II,gutierrez+17I,dejaeger+18}.
The diversity of spectroscopic properties is also observed in the near-infrared (NIR), although in this regime there may be spectroscopic differences between slow- and fast-declining \sneii\ \citep{davis+19}. 
Early-time LCs ($\lesssim$\,30 days from explosion) also contribute to \snii\ diversity \citep[e.g.][]{gonzalez+15,gall+15,ganot+20}. 
Early LCs are mostly sensitive to the characteristics of the outer envelope. The measured rise times, that is, the time from the explosion epoch to the date of maximum light, are much shorter than model predictions indicating either small and dense pre-SN envelopes, or a delayed and prolonged shock breakout because of interaction with an extended atmosphere or a shell of dense circumstellar material (CSM)  \citep{gonzalez+15,forster+18,morozova+18}. 

The observed diversity of \sneii\ may be attributed to differences in stellar evolution processes that produce progenitors with different pre-SN properties (hydrogen-rich envelope mass, progenitor size, chemical abundances, CSM properties) and/or differences in the properties of the explosion (e.g. energy deposited by the shock and \Ni\ mass).
While the underlying physical processes involved in \sneii\ are generally well understood, significant gaps remain in our understanding of how massive stars evolve to produce these type of hydrogen-rich events, particularly in regard to the different mass-loss processes involved (stellar winds, eruptions, and mass transfer in binary systems). Therefore, it is important to constrain the predominant physical properties that cause the observed diversity, and to determine the full range of parameters that produce \snii\ events.

Theoretical works have studied the diversity of \sneii\ in terms of physical properties and found that changes in the hydrogen-rich envelope mass, progenitor radius, and explosion energy produce large differences in the luminosity, duration of the optically thick phase, and expansion velocities \citep[e.g.][]{litvinova+85,young04,utrobin07,kasen+09,bersten+11,goldberg+19}.
\citet{kasen+09} showed that \Ni\ mass extends the plateau duration, although with a smaller contribution than the aforementioned physical parameters \citep[see also][]{bersten13phd}, while the mixing of \Ni\ within the ejecta tends to modify the shape of the LC \citep{bersten+11,kozyreva+19}.
In addition, it has been shown that the progenitor hydrogen-rich envelope mass affects the plateau declination rate, where smaller masses produce \sneii\ with faster declining LCs \citep[e.g.][]{bartunov+92,blinnikov+93}.
 \citet{dessart+13} presented synthetic multi-band LCs and spectra varying progenitor and explosion properties ---such as the hydrogen-rich envelope mass, explosion energy, radius, and metallicity---  that support previous findings with respect to the changes in LCs \citep[see also][]{hillier+19}. Additionally, \citet{dessart+13} showed that the progenitor radius strongly influences early colours.
The hydrogen-rich envelope mass and the explosion energy seem to be the physical parameters that most influence \snii\ LC diversity.

In this, the third paper of a series where we analyse a statistically significant sample of \sneii, we focus on understanding their observed diversity in terms of progenitor and explosion properties. 
We used a much larger set of observations than previous studies, which were directly modelled using hydrodynamical simulations that explore the most important physical parameters.

The first paper of this series \citep[][hereafter Paper~I]{martinez+21a} focuses on the calculation of bolometric LCs for the \sneii\ observed by the Carnegie Supernova Project-I \citep[CSP-I,][]{hamuy+06}. 
We found that NIR observations are crucial for accurate bolometric luminosity calculations, and that the absence of these data produces significantly different bolometric LCs. These differences are then transferred to the physical parameters derived from LC and expansion velocity modelling.
\citetalias{martinez+21a} provides relations to address the absence of NIR data, in addition to new prescriptions for bolometric corrections as a function of optical colours to be used by the community.

The bolometric LCs presented in \citetalias{martinez+21a}, together with the \ion{Fe}{ii} $\lambda$5169 line velocities measured by \citet{gutierrez+17I}, were then used for deriving progenitor and explosion properties via hydrodynamical modelling in \citet[][hereafter Paper~II]{martinez+21b}, where a large set of explosion models were used that were previously presented in \citet[][hereafter M20]{martinez+20}.
In \citetalias{martinez+21b} we also analysed the distributions of the inferred physical parameters.
The modelling of the initial-mass (\mzams) distribution gives an upper mass of $<$\,21.5~\ms, consistent with the existence of the RSG problem \citep{smartt+09}, especially when the power-law slope of the \mzams\ distribution is constrained to be that of a Salpeter massive-star initial-mass function (IMF).
However, a much steeper distribution than that for a Salpeter IMF is favoured by our modelling. We named this result `the IMF incompatibility'. This incompatibility is due to the large number of progenitors compatible with low-ejecta-mass models, possibly implying that massive stars lose more mass during their lives than predicted by standard single-star evolution, although a more thorough analysis of all stellar evolution assumptions is required to test this hypothesis.

As part of the studies presented in \citetalias{martinez+21a} and \citetalias{martinez+21b}, we built the most homogeneous and largest sample of \snii\ bolometric LCs to date for which the physical properties of the objects are determined via hydrodynamical modelling.
Here, in Paper~III, we present an analysis of correlations between observed and physical properties in order to further our understanding of \snii\ diversity in terms of the physics of the explosions and their progenitors.
The paper is organised as follows. Section~\ref{sec:sample} briefly describes the observational and modelling samples. Section~\ref{sec:results} presents the analysis of correlations when physical properties are derived using pre-SN models assuming non-rotating single-star evolution, while Sect.~\ref{sec:nonstd} shows the results when non-standard stellar evolution is considered.
In Sect.~\ref{sec:discussion} we discuss the most interesting of the correlations in detail, and present our conclusions in Sect.~\ref{sec:conclusions}.
Figures not included in the main body of the manuscript are presented in the Appendices.

%------------------------------------------------------
\section{Observational and theoretical sample}
\label{sec:sample}

%------------------------------------------------------
\subsection{Supernova sample}

The sample of \sneii\ used in this work is the same as that analysed in \citetalias{martinez+21a} and \citetalias{martinez+21b}.
The dataset was observed by the CSP-I and contains 74~\sneii. 
The CSP-I was a follow-up programme of SNe based at the Las Campanas Observatory that observed between 2004 and 2009 \citep{hamuy+06}. The CSP-I sample consists of optical ($uBgVri$) and NIR ($YJH$) LCs and optical spectra for most objects, resulting from high-cadence and high-quality observations.
CSP-I $V$-band photometry was published by \citet[][A14 hereafter]{anderson+14_lc} while the optical spectra were published by \citet{gutierrez+17II,gutierrez+17I}. The final photometry for the CSP-I \snii\ sample is presented in Anderson et al. (in prep.).
Details of these \sneii\ are available in the above-mentioned studies and \citetalias{martinez+21a}.

%------------------------------------------------------
\subsection{Observed properties}
\label{sec:measurements}

Previous studies used the \sneii\ in the CSP-I sample to analyse their spectral and photometric diversity.
\citetalias{anderson+14_lc} analysed the $V$-band LC properties through the measurement of magnitudes at different epochs, declination rates, and durations of different phases.
\citet{anderson+14_ha} and \citet{gutierrez+14} presented studies of the \ha\ profiles, specifically of the blueshifted emission-peak offset, velocity, and ratio of absorption to emission.
A posterior analysis was presented by \citet{gutierrez+17I} and \citet[][hereafter G17]{gutierrez+17II}, where the authors studied expansion velocities and pseudo-equivalent widths (pEWs) of numerous spectral lines, together with additional $V$-band LC and spectral properties.
CSP-I \snii\ colour curves were studied by \citet[][hereafter dJ18]{dejaeger+18}.

The number of photometric and spectroscopic parameters studied in the literature for the CSP-I \snii\ sample is extremely large, and a complete analysis of all previously mentioned parameters would be too long for one publication.
Therefore, in this section, we summarise the observed parameters used in the present study.
These are chosen to elucidate previous questions posed in the literature with respect to how observed parameters are determined from progenitor and/or explosion properties. 

Progenitor and explosions properties for our \snii\ sample were derived via hydrodynamical modelling of their bolometric LCs and photospheric velocities \citepalias{martinez+21b}.
Therefore, we used bolometric LC parameters for the analysis of correlations, instead of the $V$-band LC parameters measured by \citetalias{anderson+14_lc} and \citetalias{gutierrez+17II}.
\snii\ bolometric LC parameters for the CSP-I sample were measured and analysed in \citetalias{martinez+21a}, to which we refer the reader for details. We briefly outline the parameters used in the present work \citepalias[see also Fig.~8 from][]{martinez+21a}:
(1) \mbolend\ is the bolometric magnitude measured 30 days before the mid-point of the transition from plateau to the radioactive tail (\tpt); 
(2) \mboltail\ is the bolometric magnitude measured 30~days after \tpt; 
(3) \sone\ is the declination rate in magnitudes per 100~days of the cooling phase; 
(4) \stwo\ is the declination rate in magnitudes per 100~days of the plateau phase (which is not necessarily a phase of constant magnitude); 
(5) \sthr\ is the declination rate in magnitudes per 100~days of the slope in the radioactive tail phase; 
(6) \optd\ is the duration of the optically thick phase and it is equal to \tpt; 
(7) \pd\ is the duration of the plateau phase and it is equal to \tpt~$-$~\ttrans\footnote{\ttrans\ is the epoch of transition between \sone\ and \stwo.}; and 
(8) \cd\ is the duration of the cooling phase, defined between the explosion epoch and \ttrans.
In addition, we used ten spectral parameters from \citetalias{gutierrez+17II}, all measured at 50~days post-explosion:
(1) expansion velocity ($v$) for the absorption component of \ha, \hb, and \ion{Fe}{ii} $\lambda$5169 lines; 
(2) pEW of \ha\ (absorption and emission components), \hb, \ion{Fe}{ii} $\lambda$4924, \ion{Fe}{ii} $\lambda$5018, and \ion{Fe}{ii} $\lambda$5169; and
(3) flux ratio of the absorption to emission component of \ha\ P-Cygni profile (\aeha).
Colour information was also included. We used ($g-r$) colours measured at 15 and 70~days post-explosion \citepalias{dejaeger+18}.

%------------------------------------------------------
\subsection{Progenitor and explosion models}
\label{sec:models}

\input{physical_pars.tab}

The LCs and expansion velocities of \sneii\ are sensitive to the physical properties of the progenitor star and the explosion, such as the ejecta mass (\mej: pre-SN mass minus the mass of the forming compact remnant), hydrogen-rich envelope mass (\mhenv), progenitor radius (\ra), explosion energy (\e), \Ni\ mass (\mni), and its distribution within the ejecta (\mix).
In \citetalias{martinez+21b}, we used a large grid of bolometric LC and photospheric velocity models applied to stellar evolution progenitors presented in \citetalias{martinez+20} for deriving the    physical properties of the CSP-I \snii\ sample.
Here, a summary of the models is presented, but we refer the reader to \citetalias{martinez+20} for additional information.

Non-rotating solar-metallicity pre-SN RSG models were calculated for \mzams\ between 9 and 25~\ms\ with increments of 1~\ms \ using the public stellar evolution code \texttt{MESA}\footnote{http://mesa.sourceforge.net/} version 10398 \citep{paxton+11,paxton+13,paxton+15,paxton+18,paxton+19}. The stellar models were evolved from the pre-main sequence until core collapse, except for the 9, 10, and 11~\ms\ progenitor models that were calculated up to core carbon depletion, because more advanced burning phases are computationally expensive.
Stellar evolution was followed using the standard prescriptions for mass loss and convection. 
For convection, the Ledoux criterion was adopted with a mixing-length parameter of 2.0. The wind mass loss was calculated from the `Dutch' prescription defined in the \texttt{MESA} code with an efficiency of $\eta$\,=\,1 \citep{dejager+88,vink+01,glebbeek+09}.
Hydrogen-rich envelope masses for this set of progenitor models cover the range of 7.1$-$10.4~\ms, while progenitor radii are found in the range of 445$-$1085~\rs.
These pre-SN stellar models were used as input to the 1D Lagrangian hydrodynamical code presented in \citet{bersten+11} to compute bolometric LCs and photospheric velocities of \sneii. The grid of explosion models covers a wide range of explosion parameters (\e, \mni, and \mix), which is given in Table~\ref{tab:models}.
We refer to these models as `standard models'\footnote{The grid of bolometric LCs and photospheric velocity models can be downloaded from \url{https://doi.org/10.5281/zenodo.6228795}.}.
The explosion models were calculated without including any CSM shell surrounding the progenitor star.

In \citetalias{martinez+21b}, the determination of the physical properties of the \sneii\ in the CSP-I sample is based on describing the bolometric LC and photospheric velocities  simultaneously by means of comparing hydrodynamical models with observations.
The bolometric LCs for the CSP-I sample were presented in \citetalias{martinez+21a}. \ion{Fe}{ii} line velocities measured by \citet{gutierrez+17I} were used, assuming this line as a proxy for the photospheric velocity.

The posterior probability distributions of the physical parameters (\mzams, \e, \mni, and \mix) for the \sneii\ in the CSP-I sample were found by employing Markov chain Monte Carlo (MCMC) methods via the \texttt{emcee} package \citep{emcee} following \citet{forster+18}.
For this, the interpolation method presented in \citet{forster+18} was used to quickly interpolate between the set of hydrodynamical models described above.
The MCMC sampler assumes flat distributions as priors for the four physical parameters of our model (\mzams, \e, \mni, and \mix) and also for the explosion epoch (\texp). The sampler was allowed to run within the ranges of the physical parameters in our set of hydrodynamical models (Table~\ref{tab:models}) and within the observational error of \texp\ \citepalias[see][]{martinez+21a}.
An additional parameter was defined, the `scale', which multiplies the bolometric luminosity by a constant dimensionless factor to allow for errors in the bolometric LC introduced by the uncertainties in the distance and host-galaxy extinction. A Gaussian prior was used for the scale \citepalias[see][for further details]{martinez+21b}.
Thus, in the current work, the physical parameters determined in \citetalias{martinez+21b} were used: \mzams, \e, \mni, and \mix.
Additionally, given that we used stellar evolution calculations as pre-SN models, each \mzams\ relates to a pre-SN structure, for which \mej, \ra, and \mhenv\ have been determined. Therefore, we also used the last three parameters to analyse possible correlations.
The errors on our estimated physical parameters are statistical in nature, and do not take into account systematic errors such as the uncertainties in stellar evolution modelling. As a consequence, the errors on the progenitor parameters are likely to be underestimated. The size of the errors indicates the high quality of the observations and the robustness of our fitting technique.

Together with the observed parameters mentioned in Sect.~\ref{sec:measurements}, we also measured observables from our extensive grid of hydrodynamical LC and photospheric velocity models for determining the effect of each physical parameter with respect to the observed \snii\ diversity.
We used the same grid of explosion models previously mentioned, that is, our standard grid of models.
While this grid was built covering a wide range of progenitor and explosion parameters, we used the interpolation technique presented in \citet{forster+18} to obtain an even larger number of measurements.
The generated grid of models covers the following parameter space: \mzams\ between 9 and 25~\ms\ in steps of 1~\ms\ and explosion energies between 0.1 and 1.5~foe (1~foe~$\equiv$~10$^{51}$~erg) in intervals of 0.05~foe, with the exception of the largest masses and lower explosion energies, because these models could not be calculated for numerical problems \citepalias[see][for details]{martinez+20}. \mni\ ranges between 0.01 and 0.08~\ms\ in steps of 0.005~\ms\ with a degree of mixing of between 0.2 and 0.8 (given as a fraction of the pre-SN mass) in intervals of 0.1.
For each model, we measured \optd, \mbolend, \mboltail, \stwo, and \sthr\ using the same definitions as in Sect.~\ref{sec:measurements}. Additionally, we measured the photospheric velocity at 50~days from explosion (\vphfi).

Models were constructed covering regular ranges of physical parameters, but a subset of \snii\ models present bolometric LC parameters that have not (yet) been observed in nature, although they are theoretically possible.
Some of the bolometric LC models for low \e\ and relatively high \mhenv\ yield \optd\ values that are larger than any \sneii\ observed to date. For this reason, we only analysed models with \optd\ shorter than 160~days. This criterion is somewhat arbitrary but it is $\sim$15~days longer than the longest \snii\ plateau observed to date\footnote{This choice does not affect the conclusions of this study, but we prefer to remove these models to understand the \sneii\ in our sample.}: SN~2009ib \citep{takats+15}.
In total, 38757 measurements of each observational parameter are available.

We did not measure \pd\ in our models for the following reason: according to the parameter definitions in Sect.~\ref{sec:measurements}, \pd\ needs previous measurements of two parameters: \tpt\ (\optd) and \ttrans. The latter parameter is defined as the epoch of transition between \sone\ and \stwo, although it can also be understood as the epoch of transition between the cooling and plateau phases (based on the definitions from Sect.~\ref{sec:measurements}).
However, it is usually found that the observed cooling phase is longer than model predictions \citep[and rise times to maximum light of optical LCs are shorter, e.g.][]{gonzalez+15} for which the presence of additional material confined close to the progenitor star has been suggested as an explanation.
Our explosion models were calculated without the presence of any possible circumstellar material.
Therefore, the analysis of model \cd\ and \pd\ would be biased to smaller and larger values, respectively.

%------------------------------------------------------
\section{Results}
\label{sec:results}

\input{kda.tab}

In this section, we search for correlations between the observed and physical parameters described in Sects.~\ref{sec:measurements} and \ref{sec:models}.
Physical parameters for the CSP-I \snii\ sample were inferred in \citetalias{martinez+21b}, where the results were classified into two groups: the `gold' and `full' samples.
\sneii\ with extensive data coverage that are well reproduced by our models were classified as `gold events' (24 objects), but in total we inferred physical parameters for 53 objects \citepalias[see][for details]{martinez+21b}. In the subsequent figures, we label these distinct samples with different symbols and colours.

Throughout the rest of the paper, we used the Pearson test on the full sample of \sneii\ to determine the existence and strength of correlations by employing 10000 Monte Carlo bootstrapping simulations. For each simulation, $N$ random values were drawn allowing multiple events to be taken (where $N$ is the number of events for each correlation), and the Pearson correlation coefficient was calculated.
The distribution of the correlation coefficient is symmetric. Therefore, the mean correlation coefficient (\p) of these 10000 simulations and the standard deviation ($\sigma$) were used to characterise the correlations. These values are presented in each figure. In addition, an upper limit to the probability of finding such a correlation strength by chance ($P$) is presented.
The gold sample was also analysed, showing similar correlation coefficients to those of the full sample. In most cases, the correlation coefficients estimated for the gold sample are within the error bars of the coefficients for the full sample. Only a small number of trends show notably different correlations. These are the trends involving $(g-r)$ colours at 15~days post-explosion, because of the low number of events in the gold sample for which this parameter has been measured. For these reasons, we only present our analysis of correlations for the full sample.
We used the following descriptions to characterise the strength of correlations: correlation coefficients between 0 and 0.19 show zero correlation, 0.20$-$0.39 weak, 0.40$-$0.59 moderate, 0.60$-$0.79 strong, and 0.8$-$1.0 very strong \citep{evans96}.

The observables measured from the grid of explosion models allow us to perform a statistical analysis of the influence of each individual physical parameter on the observed diversity of \sneii.
We performed a key driver analysis (KDA) using the \texttt{python} library \texttt{Kruskals}\footnote{\url{https://github.com/Rambatino/Kruskals}} to determine the effect of physical properties on the observed parameters \citep{kruskal87}.
KDA is a technique used to identify which of a set of independent variables causes the largest impact on a given dependent variable. 
Table~\ref{table:kda} reports the relative importance of the physical parameters (\mzams, \e, \mni, and \mix) for each parameter measured from our hydrodynamical models.
We used \mzams\ given that it is the independent variable related to a unique pre-SN structure (in the context of standard single-star evolution) emphasising that \mzams\ represents the effect of \mej\ and \ra\ simultaneously, and that they cannot be separated given their dependency on \mzams.
We note that our results are relevant for the standard stellar evolution adopted in this study and the inclusion of additional pre-SN models could modify the relative effect of each physical parameter on the observables (see Sect.~\ref{sec:corr_nonstd}).
A description of the results from Table~\ref{table:kda} is found in the following section together with the analysis of correlations using the CSP-I \snii\ sample. In addition, Appendix~\ref{appendix:plots} includes figures that show observables measured from the models against the physical parameter yielding the highest relative importance, and a supplementary analysis of the relations found.

We separate the analysis of correlations in the following subsections. Correlations between observed and physical parameters for the CSP-I \snii\ sample are presented in Sect.~\ref{sec:corr_std}. In Sect.~\ref{sec:corr_physical_pars}, we then look for correlations between physical parameters  only.
In addition, an Appendix is included where the main trends between different observed spectral, colour, and bolometric LC parameters are presented\footnote{We present those in the Appendix given that most of that work duplicates previous efforts using this same sample. The only main difference is that here we use bolometric LC parameters in place of the $V$ band.} (Appendix~\ref{sec:corr_obs_pars}). The reader is referred to those pages for a complete analysis of the correlations.
In Sect.~\ref{sec:nonstd} we add pre-SN models evolved with an enhanced mass-loss rate to our grid of explosion simulations to fit some \sneii\ that are not well reproduced by standard single-star models. Additionally, these models are included in the fitting procedure to the full sample and correlations are re-analysed. Some of the correlations increase in strength when including models with enhanced mass loss.

%------------------------------------------------------
\subsection{Correlations between physical and observed parameters using 53~\sneii\ from the CSP-I sample}
\label{sec:corr_std}

\begin{figure*}
\centering
\includegraphics[width=1.0\textwidth]{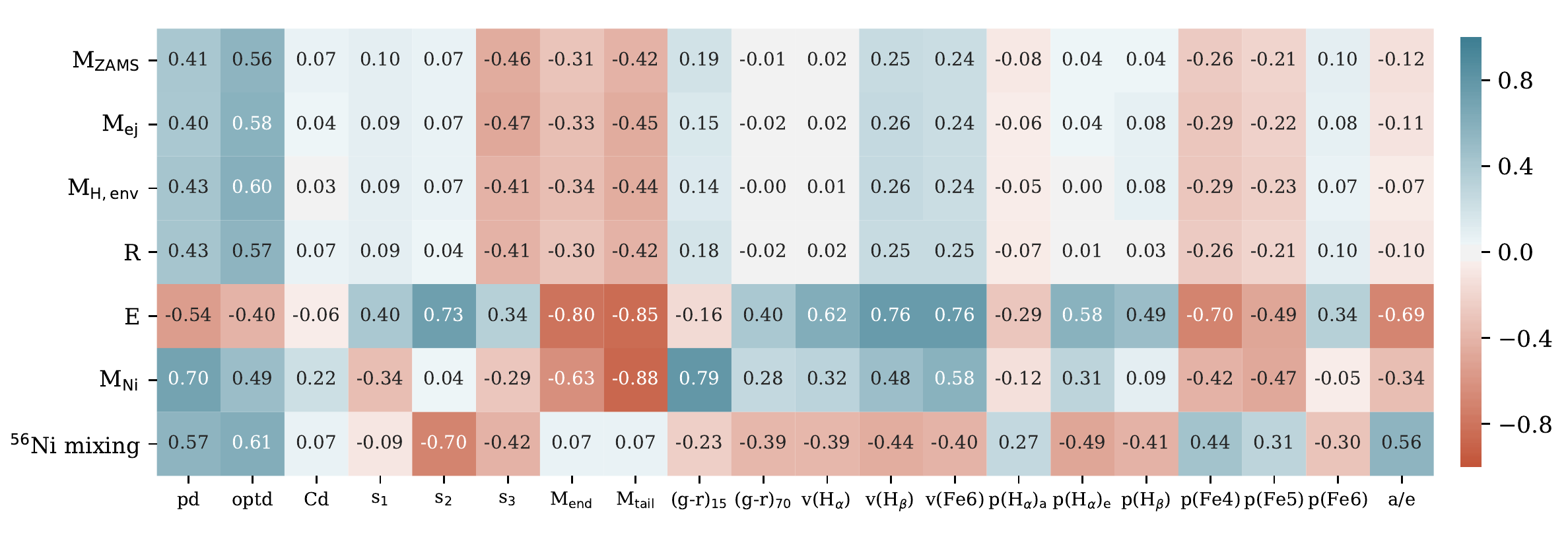}
\caption{Correlation matrix between observed and physical parameters of \sneii. For each pair, the Pearson correlation coefficient is given and is colour-coded. The observed parameters shown are: \pd, \optd, \cd, \sone, \stwo, \sthr, \mbolend, \mboltail, \gr$_{15}$, \gr$_{70}$, velocities of \ha, \hb, and \ion{Fe}{ii} $\lambda$5169, pEW(\ha) of absorption component, pEW(\ha) of emission component, pEW(\hb), pEW(\ion{Fe}{ii} $\lambda$4924), pEW(\ion{Fe}{ii} $\lambda$5018), pEW(\ion{Fe}{ii} $\lambda$5169), and \aeha.}
\label{fig:corr_matrix_physical_observed_std}
\end{figure*}

\begin{figure*}
\centering
\includegraphics[width=1.0\textwidth]{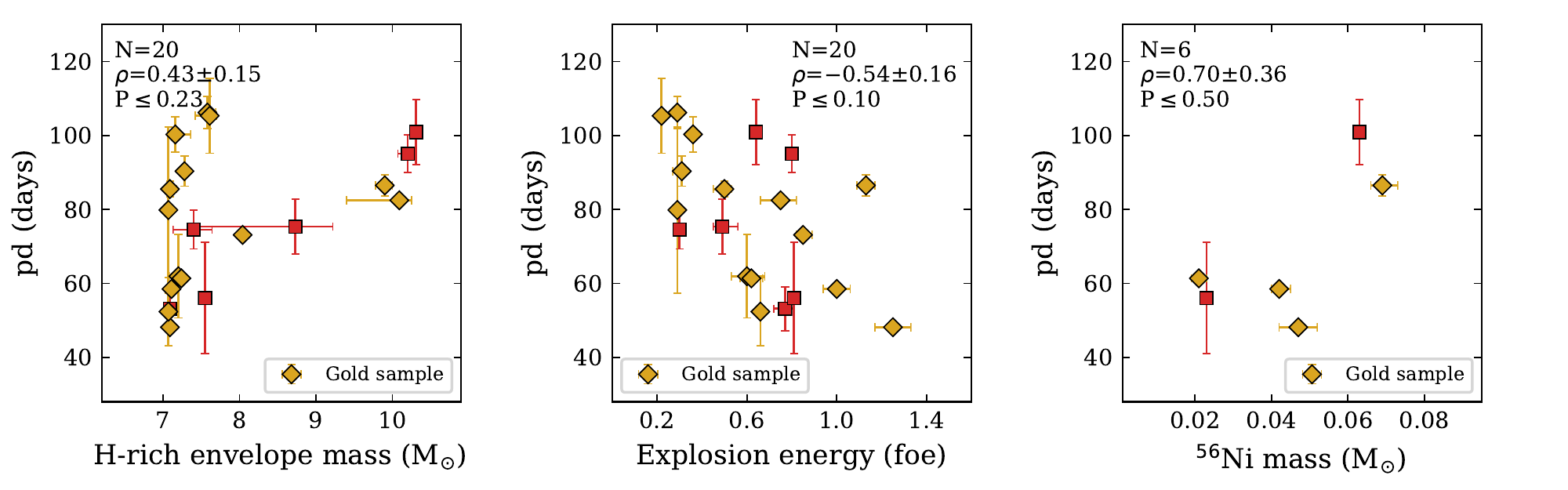}
\caption{Correlations between \pd\ and three physical parameters: \mhenv\ (left panel), \e\ (middle panel), and \mni\ (right panel). The yellow and red markers refer to the results obtained from the CSP-I \snii\ sample (yellow markers indicate gold events). Each subplot contains the number of events ($N$), the Pearson correlation coefficient (\p), and the probability of detecting a correlation by chance ($P$) using the full sample.}
\label{fig:corr_pd_std}
\end{figure*}

\begin{figure*}
\centering
\includegraphics[width=0.74\textwidth]{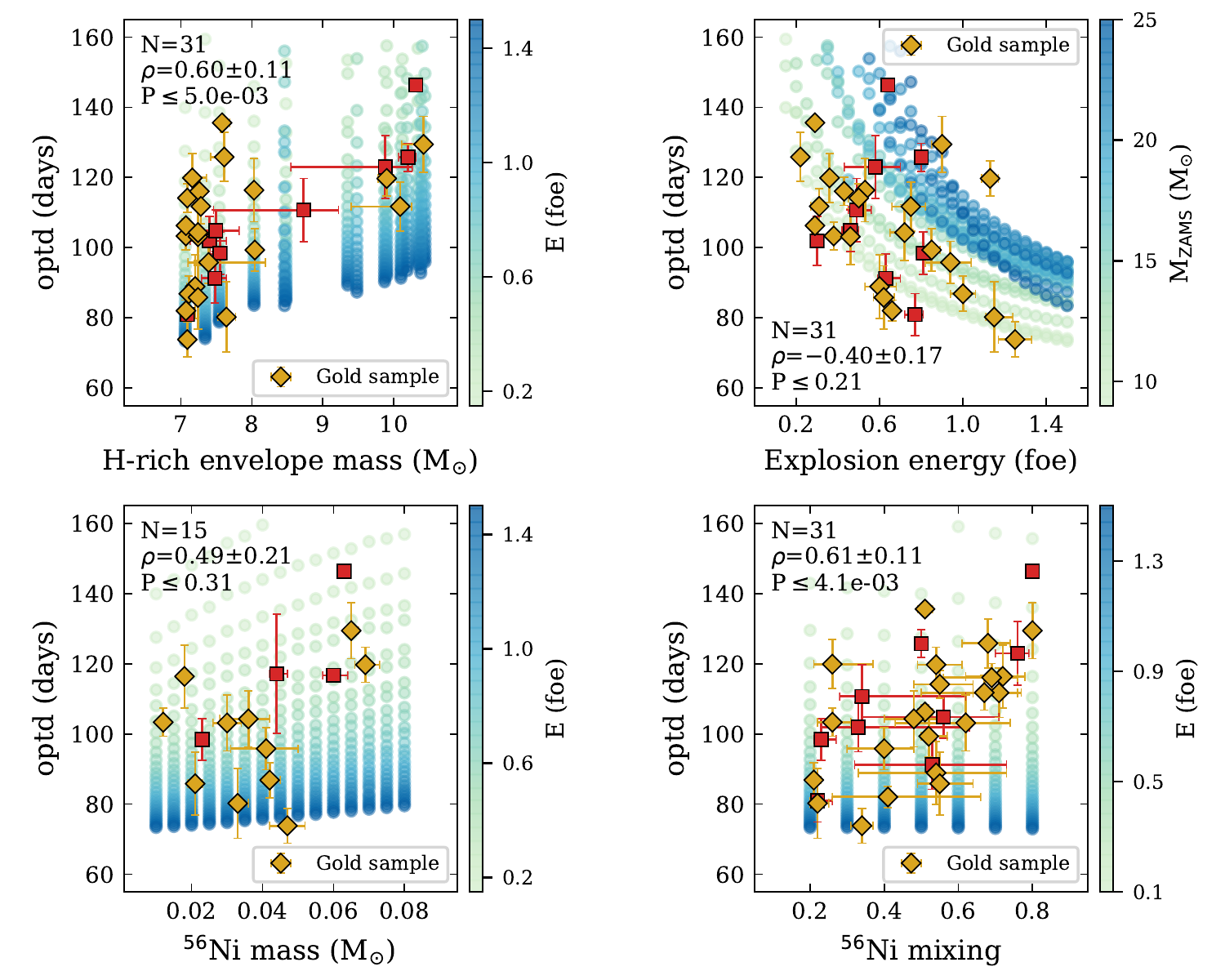}
\caption{Correlations between \optd\ and physical parameters: \mhenv\ (top left), \e\ (top right), \mni\ (bottom left), and \mix\ (bottom right). The yellow and red markers refer to the results obtained from the CSP-I \snii\ sample (yellow markers indicate gold events). Each subplot contains the number of events ($N$), the Pearson correlation coefficient (\p), and the probability of detecting a correlation by chance ($P$) using the full sample. Results from the models are colour-coded based on different physical parameters. The physical parameters not being varied ---if they do not appear in the plot--- are fixed at \mzams\,=\,10~\ms, \mni\,=\,0.01~\ms, and \mix\,=\,0.5. Some observations fall outside the range of the model parameters because of the fixed physical parameters. Changes in the fixed values represent different ranges of model parameters.}
\label{fig:corr_optd_std}
\end{figure*}

\begin{figure*}
\centering
\includegraphics[width=0.74\textwidth]{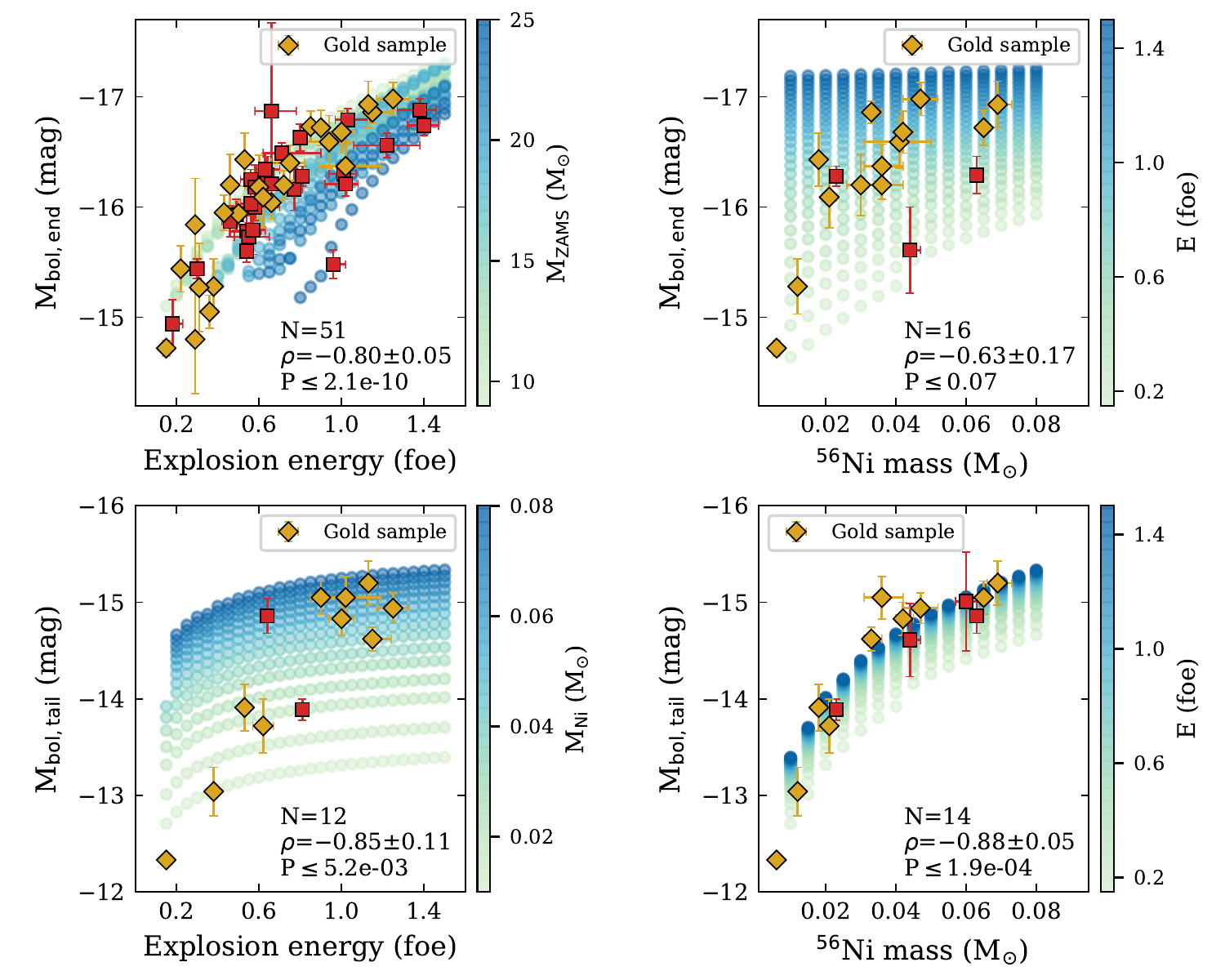}
\caption{Correlations between physical parameters and magnitude at different epochs. The top panels show correlations between \mbolend\ and two physical parameters: \e\ (top left) and \mni\ (top right). The bottom panels show correlations between \mboltail\ and two physical parameters: \e\ (bottom left) and \mni\ (bottom right). The yellow and red markers refer to the results obtained from the CSP-I \snii\ sample (yellow markers indicate gold events). Each subplot contains the number of events ($N$), the Pearson correlation coefficient (\p), and the probability of detecting a correlation by chance ($P$) for the full sample. Results from the models are colour-coded based on different physical parameters. The physical parameters not being varied ---if they do not appear in the plot--- are fixed at \mzams\,=\,10~\ms, \mni\,=\,0.03~\ms, and \mix\,=\,0.5. Some observations fall outside the range of the model parameters because of the fixed physical parameters. Changes in the fixed values represent different ranges of model parameters.}
\label{fig:corr_mbol_std}
\end{figure*}

\begin{figure*}
\centering
\includegraphics[width=1.0\textwidth]{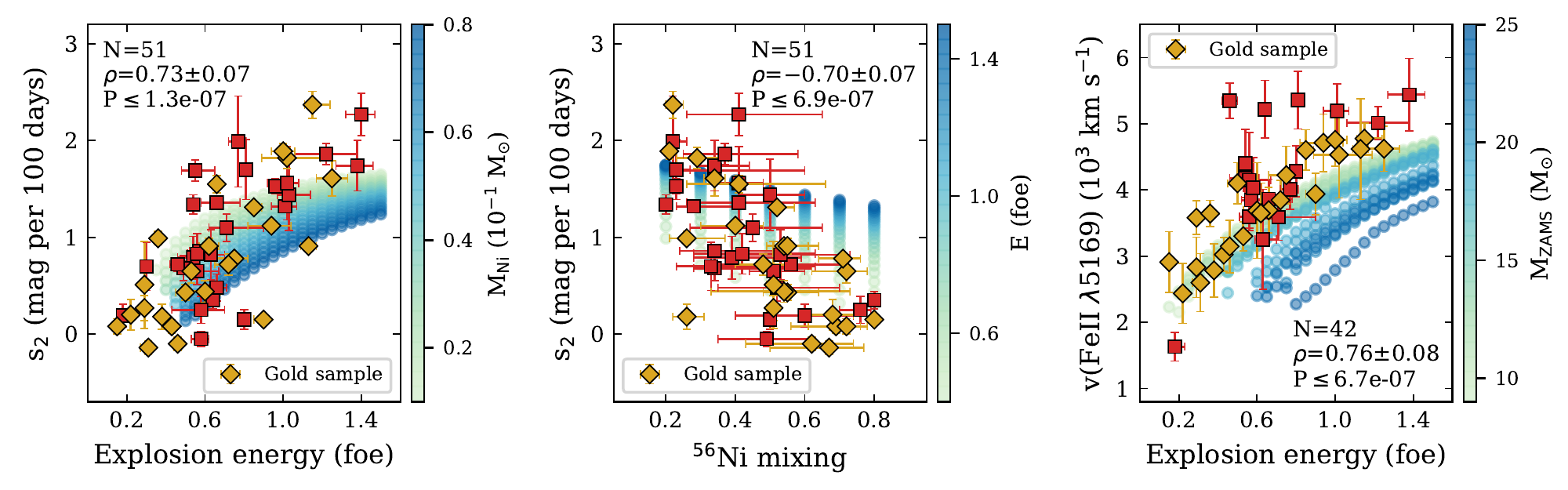}
\caption{Correlations between \stwo\ and \e\ (left panel), \stwo\ and \mix\ (middle panel), and \ion{Fe}{ii} $\lambda$5169 line velocities at 50~days post-explosion and \e\ (right panel). The mixing of \Ni\ within the ejecta is given as a fraction of the pre-SN mass. The yellow and red markers refer to the results obtained from the CSP-I \snii\ sample (yellow markers indicate gold events). Each subplot contains the number of events ($N$), the Pearson correlation coefficient (\p), and the probability of detecting a correlation by chance ($P$) using the full sample. Results from the models are colour-coded based on different physical parameters. The physical parameters not being varied ---if they do not appear in the plot--- are fixed at \mzams\,=\,15~\ms, \mni\,=\,0.04~\ms, and \mix\,=\,0.5. Some observations fall outside the range of the model parameters because of the fixed physical parameters. Changes in the fixed values represent different ranges of model parameters.}
\label{fig:corr_s2_vel_std}
\end{figure*}

\begin{figure*}
\centering
\includegraphics[width=1.0\textwidth]{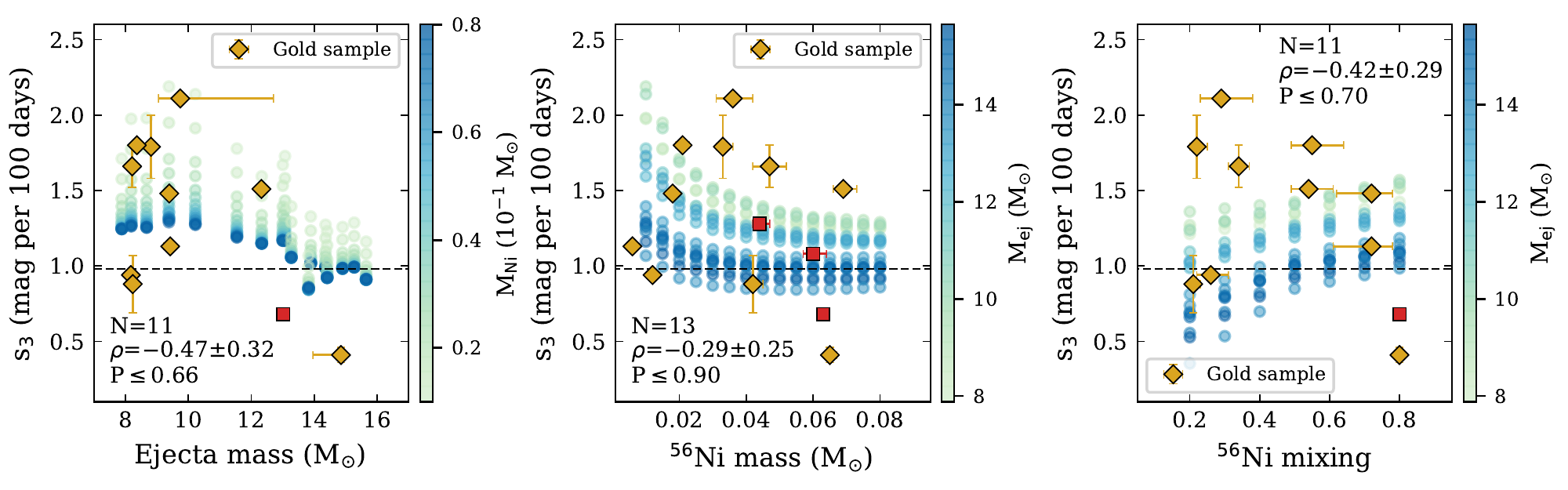}
\caption{Correlations between \sthr\ and three physical parameters: \mej\ (left panel), \mni\ (middle panel), and \mix\ (right panel). The dashed line indicates the expected declination rate for full trapping of $^{56}$Co decay. The yellow and red markers refer to the results obtained from the CSP-I \snii\ sample (yellow markers indicate gold events). Each subplot contains the number of events ($N$), the Pearson correlation coefficient (\p), and the probability of detecting a correlation by chance ($P$) using the full sample. Results from the models are colour-coded based on different physical parameters. The physical parameters not being varied ---if they do not appear in the plot--- are fixed at \e\,=\,1.2~foe, \mni\,=\,0.04~\ms, and \mix\,=\,0.5. Some observations fall outside the range of the model parameters because of the fixed physical parameters. Changes in the fixed values represent different ranges of model parameters.}
\label{fig:corr_s3_std}
\end{figure*}

Figure~\ref{fig:corr_matrix_physical_observed_std} shows the correlation matrix for the progenitor properties and observables that we consider in this study. For each pair, the Pearson coefficient is given and colour-coded: darkest colours represent the highest correlations, while white colour indicates no correlation.
In Sect.~\ref{sec:corr_physical_pars}, we show the strong correlations between progenitor parameters (\mzams, \mej, \mhenv, and \ra) which are inherent to the progenitor models calculated assuming standard stellar evolution (i.e. they are all determined by \mzams). 
Despite this, we searched for correlations between all progenitor parameters and observables for a more appropriate comparison with previous works.

The figures presented in this section show correlations between physical and observed properties for the CSP-I \snii\ sample. Each figure also shows results from the models (if the observed parameter has been measured). These are colour-coded based on the physical parameter that produces the largest impact on the observable being analysed (adopting the results from Table~\ref{table:kda}) unless the physical parameter is already in the plot. In that case, the parameter that produces the second-largest impact is used. The other physical parameters are fixed.
Some observations fall outside the range of the model parameters, which is due to the fixed physical parameters. Changes in these values will produce different ranges of model observables.

Figure~\ref{fig:corr_pd_std} shows relations between \pd\ and three physical parameters: \mhenv, \e, and \mni. A positive trend is found between \pd\ and \mhenv\ (left panel), with a correlation coefficient of \p\,=\,0.43\,$\pm$\,0.15 ($N$\,=\,20). At the same time, \pd\ shows an evident trend with the explosion energy (\p\,=\,$-$0.54\,$\pm$\,0.16, $N$\,=\,20).
It is interesting to note that while the two \sneii\ in our sample with the longest \pd\ are consistent with the lowest-energy explosions, the \snii\ with the shortest \pd\ is the most energetic.
A trend is also found between \pd\ and \mni, although our sample size is too small (only six points) for us to make any strong conclusions.

Figure~\ref{fig:corr_optd_std} shows correlations between \optd\ and physical parameters: \mhenv\ (top-left panel), \e\ (top-right panel), \mni\ (bottom-left panel), and \mix\ (bottom-right panel). \optd\ displays a strong correlation with \mhenv\ (\p\,=\,0.60\,$\pm$\,0.11 $N$\,=\,31).
Given that this correlation coefficient is larger than that for the \pd$-$\mhenv\ relation, in addition to the higher significance, these might suggest that \optd\ is a better indicator of the hydrogen-rich envelope mass than \pd. 
This is the opposite of what is claimed by \citetalias{gutierrez+17II}. We note that different definitions are used for the `plateau duration' and `optically thick phase duration' in the two studies (\citetalias{gutierrez+17II} named these parameters as Pd and OPTd, respectively; see Sect.~\ref{sec:measurements} and \citetalias{martinez+21a}). Therefore, we also estimated the correlation coefficients between \mhenv\ and both Pd and OPTd using the observed values measured by \citetalias{gutierrez+17II}, finding again that the optically thick phase duration shows a stronger correlation with \mhenv.

In opposition to \pd, we find that \optd\ exhibits a higher degree of correlation with \mhenv\ than with \e\ (\p\,=\,$-$0.40\,$\pm$\,0.17, $N$\,=\,31). 
A trend is found with \mni, in the sense that longer \optd\ are consistent with higher \Ni\ values, although with a large dispersion. We observe a strong correlation between \optd\ and the mixing of \Ni\ within the ejecta (\p\,=\,0.61\,$\pm$\,0.11, $N$\,=\,31).
Figure \ref{fig:corr_optd_std} also displays the results measured directly from the models. In line with previous theoretical predictions, it is seen that larger \optd\ values are found for higher \mhenv\ and \mni, and lower \e. However, different \mix\ in the ejecta does not alter the duration of the optically thick phase, in contrast to the strong correlation found using the CSP-I \snii\ sample (see discussion in Sect.~\ref{sec:discussion}).

The analysis presented in Table~\ref{table:kda} shows that the explosion energy is the physical parameter that produces the largest impact on \optd, while \mzams\ (directly related to \mhenv, see Fig.~\ref{fig:corr_matrix_physical_pars}) produces the second-largest impact, similar to what is found from the correlations. As previously mentioned, our analysis is based on standard single-star evolution and models with different input physics could modify our statistical analysis.
As expected, \mni\ influences \optd, but with a lower relative importance than \e\ and \mzams. 

Figure~\ref{fig:corr_mbol_std} presents correlations involving magnitude measurements and physical properties. Very strong correlations are found between \mbolend\ and \e\ (\p\,=\,$-$0.80\,$\pm$\,0.05, $N$\,=\,51), and \mboltail\ and \e\ (\p\,=\,$-$0.85\,$\pm$\,0.11, $N$\,=\,12), with more energetic explosions producing brighter \sneii\ during the plateau and radioactive tail.
Table~\ref{table:kda} argues that \mbolend\ is mostly affected by \e, while \mzams\ is the physical parameter that produces the largest deviation in the \e$-$\mbolend\ relation.
A strong correlation is found between \mbolend\ and \mni\ (\p\,=\,$-$0.63\,$\pm$\,0.17, $N$\,=\,16). 
The additional heating of the ejecta at late times by the \Ni\ decay chain not only extends the duration of the plateau \citep[e.g.][]{kasen+09}, but it also increases the luminosity
in the late-plateau phase \citep{bersten13phd,kozyreva+19}.
Our models show that \mni\ has its major effect on \mbolend\ in the low-\e\ regime (\e\,$\lesssim$\,0.7~foe) when \sneii\ are fainter (Fig.~\ref{fig:mbolend}, middle panel). Because of the low luminosity, \Ni\ plays a more important role producing higher luminosities at the end of the plateau. Therefore, \mni\ also affects \mbolend, although other physical properties such as \e\ and \mzams\ have larger effects (Table~\ref{table:kda}).
Additionally, Fig.~\ref{fig:corr_mbol_std} (bottom-right panel) shows a very strong correlation between \mni\ and \mboltail\ (\p~=~$-$0.88~$\pm$~0.05, $N$=14). This is to be expected given that the tail luminosity is predominantly related to the amount of \Ni\ in the ejecta. 
The explosion energy and \mix\ deviate the tight correlation between \mni\ and \mboltail\ only on small scales (see Table~\ref{table:kda} and Fig.~\ref{fig:mboltail}).

The left and middle panels of Fig.~\ref{fig:corr_s2_vel_std} explore the correlations found between the \stwo\ declination rate and physical parameters.
We observe strong correlations with the explosion energy (\p\,=\,0.73\,$\pm$\,0.07, $N$\,=\,51), which suggests that more steeply declining \sneii\ during the plateau phase are produced by more energetic explosions. 
In addition, it is found that \stwo\ and \mix\ show a strong anti-correlation with fast-declining \sneii, which is\ compatible with more concentrated \Ni\ in the inner regions of the ejecta \citep{bersten+11}.
According to Table~\ref{table:kda}, the explosion energy causes the largest effect on the \stwo\ diversity, while \mix\ displays smaller changes.
The KDA finds that \mni\ also influences variations in \stwo. As stated above, higher \mni\ can boost the late-plateau luminosity producing slowly declining events (see also Fig.~\ref{fig:s2}, left panel). However, here we find zero correlation between \mni\ and \stwo\ when the CSP-I \snii\ sample is used (Fig.~\ref{fig:corr_matrix_physical_observed_std}).
Previous studies suggested that progenitors with smaller hydrogen-rich envelope masses produce faster declining \sneii\ \citep[i.e. larger \stwo; e.g.][]{blinnikov+93}. 
However, we find zero correlation between \mhenv\ and \stwo\ in the present work (see Fig.~\ref{fig:corr_hmass_s2_std}). 
The study of correlations shows that the \stwo\ diversity is mostly related to changes in \e\ and \mix, although the analysis of the KDA technique indicates that \e\ is the physical parameter that most influences variations of \stwo.
However, the analysis we are carrying out in this section only involves standard pre-SN models, where none of the progenitors were evolved with significant mass loss.
Therefore, the effect of \mhenv\ is probably underestimated in this analysis.

As expected, the explosion energy shows a strong correlation with \ion{Fe}{ii} $\lambda$5169 velocity (Fig.~\ref{fig:corr_s2_vel_std}, right panel), although there are outliers from this relation that display high \ion{Fe}{ii} velocities but lower energies than expected. These \sneii\ are SN~2004er and SN~2007sq. In \citetalias{martinez+21b}, the bolometric LCs of these two \sneii\ are generally well reproduced, but for both SNe our models underestimate their photospheric velocities. 
SN~2004er shows a moderately luminous and considerably long \optd. While a more energetic explosion would give much better agreement with the observed velocities, it would also lead to shorter \optd. This is similar to what happens for SN~2007sq (see \citetalias{martinez+21b} for discussion).
This shows that higher explosion energies and different pre-SN structures (e.g. the core to envelope mass) are required in both cases for a more appropriate modelling. Unfortunately, for both \sneii, these proposed models fall outside our grid.

The \e\ and \mzams\ are the only two physical parameters that produce different photospheric velocities during the optically thick phase, with \e\ showing the highest relative importance (Table~\ref{table:kda}).
However, a moderate correlation is found between \mix\ and the \ion{Fe}{II} $\lambda$5169 expansion velocities (Fig.~\ref{fig:corr_matrix_physical_observed_std}). 
This relation is possibly driven by other parameters that simultaneously correlate with the mixing of \Ni\ and expansion velocities.
Previously, we noted that faster declining \sneii\ are well reproduced by low degrees of \mix. At the same time, faster declining \sneii\ present higher expansion velocities \citepalias[e.g.][]{gutierrez+17II}. Thus, this combination of correlations may produce the trend between \mix\ and \ion{Fe}{II} $\lambda$5169 velocities.

The declination rate of the radioactive tail phase (\sthr) takes the theoretical value of 0.98~mag per 100 days assuming full trapping of the gamma-ray emission from $^{56}$Co decay (dashed lines in Fig.~\ref{fig:corr_s3_std}). The full trapping may be possible by a large ejecta mass, which results in a long diffusion time.
Observations show higher \sthr\ for a considerable number of \sneii\ \citepalias{anderson+14_lc,gutierrez+17II,martinez+21a}.
Less trapping of gamma rays as a result of low ejecta mass and/or low density can explain the observed higher \sthr\ values.
In Fig.~\ref{fig:corr_s3_std}, the \sthr\ declination rate is plotted against three physical parameters. The \sthr\ shows moderate trends with \mej\ (left panel) and \mix\ (right panel). 
In addition, a weak trend is detected with \mni\ (middle panel).  However, in all of these cases, the number of events is low and the scatter is large, which prevents us from making definitive conclusions.
Some of these relations are reanalysed in Sect.~\ref{sec:corr_nonstd}.
For a small number of \sneii,\ the luminosity decays slower than what is assumed for full trapping (i.e. lower \sthr\ values). This may be caused by the presence of residual thermal energy of the explosion in the ejecta \citep{utrobin07}, interaction with CSM \citep{fraser+15,pastorello+18}, late-time accretion onto a compact remnant \citep[e.g.][]{moriya+18,gutierrez+20}, or by a more quickly receding photosphere \citep[e.g. increasing helium abundances,][]{chieffi+03}, in addition to the contribution of $^{56}$Co~decay.

The analysis of the model parameters shows that \sthr\ is mostly affected by \mzams\ (Table~\ref{table:kda}, see also Fig.~\ref{fig:s3}). A more massive star implies larger \mej\ and \ra, at least within the range of initial masses of our standard pre-SN models. Therefore, the high \sthr\ values are consistent with the lowest ejecta masses in our grid of models.
Our models show that the degree of \Ni\ mixing within the ejecta also contributes to the differences found in \sthr\ (Table~\ref{table:kda}).
Extensively distributed \Ni\ in the envelope allows gamma-ray photons emitted from the decay of the $^{56}$Co in the outermost layers of the ejecta to easily escape before being thermalised due to the low mass beyond the location where they are emitted.
This is seen in the right panel of Fig.~\ref{fig:corr_s3_std} for the observables measured from the models, where more extended \mix\ produces \sneii\ declining more rapidly during the radioactive tail. However, this is in tension with the relation found using the \sthr\ declination rate from the observations.
Extended mixing of \Ni\ is necessary to fit some aspects of the observations, although this does not explain the large \sthr\ values measured. This might suggest that the \mix\ is not the main driver of the observed diversity of \sthr\ declination rates.

We do not find correlations between physical properties and the early-time LC parameters \cd\ and \sone\ (with the exception of \sone\ and \e, see Fig.~\ref{fig:corr_matrix_physical_observed_std}). This is not surprising because a large fraction of \sneii\ may experience interaction between the SN ejecta and a dense CSM shell surrounding the progenitor star \citep{gonzalez+15,forster+18,bruch+21}. Thus, early phases of SN evolution would be sensitive to the CSM characteristics, which are not studied in the present work.

The \gr\ colour at 15~days from explosion shows a trend with \mni, although only five points are available to compute the correlation strength. At 70~days from explosion, the \gr\ colour moderately correlates with the explosion energy, in the sense that more energetic \sneii\ are redder at late times. We further analyse this correlation in Sect.~\ref{sec:discussion}.

Finally, correlations between \e\ and the pEWs of some lines are mentioned. Figure~\ref{fig:corr_matrix_physical_observed_std} shows an evident trend with \aeha\ (\p\,=\,$-$0.69\,$\pm$\,0.07) in the sense that more energetic explosions display lower \aeha\ values. 
We also find that \e\ anti-correlates with the pEWs of \ion{Fe}{ii} $\lambda$4924 and \ion{Fe}{ii} $\lambda$5018 (\p\,=\,$-$0.70\,$\pm$\,0.08 and \p\,=\,$-$0.49\,$\pm$\,0.14, respectively), although positive correlations are found with the pEWs of \ion{Fe}{ii} $\lambda$5169, the emission component of \ha, and the absorption component of \hb.
None of the spectral and colour parameters show any significant correlation with those physical parameters dependent on \mzams.

%------------------------------------------------------
\subsection{Correlations between physical parameters}
\label{sec:corr_physical_pars}

\begin{figure}
\centering
\includegraphics[width=0.49\textwidth]{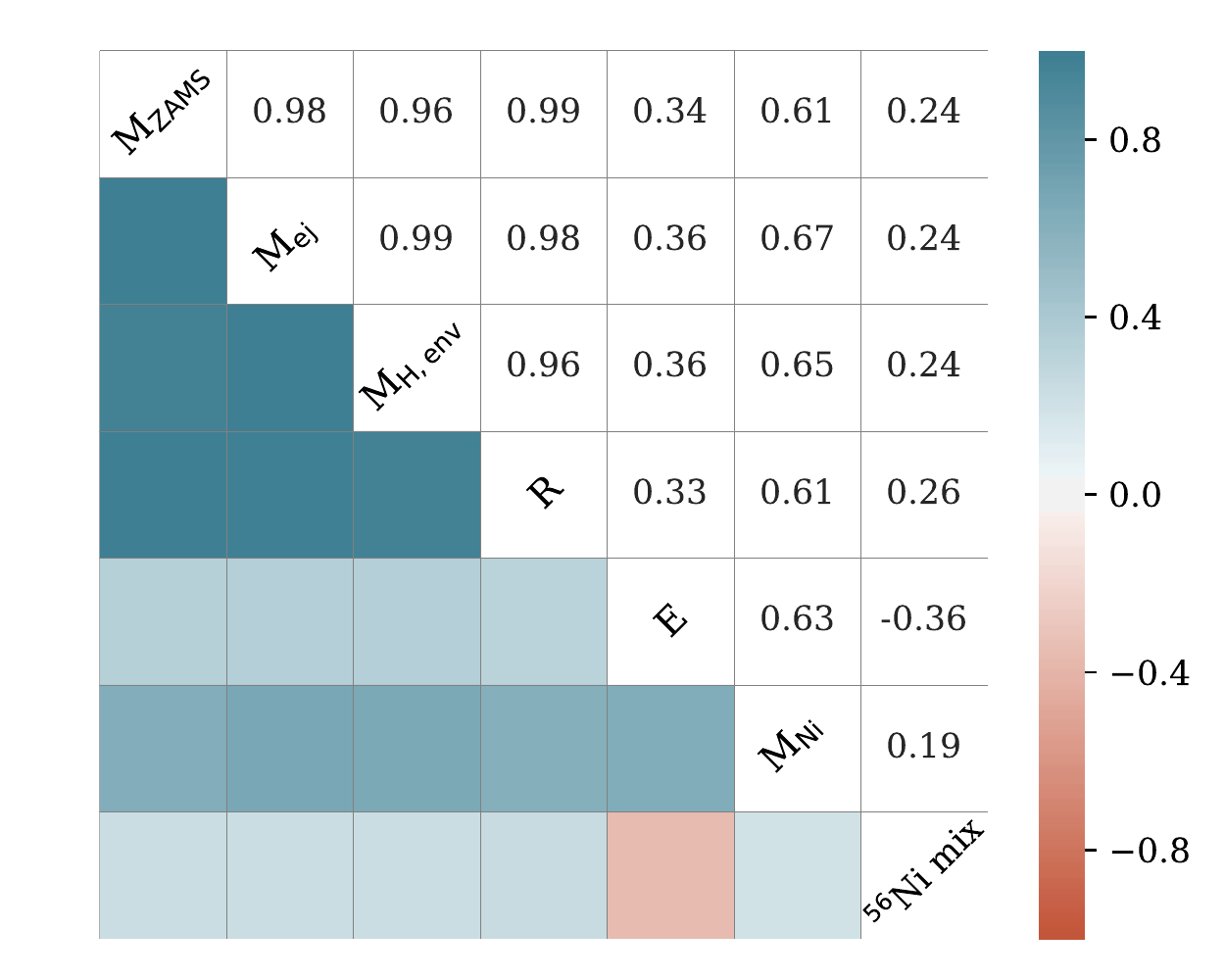}
\caption{Correlation matrix of the progenitor and explosion parameters. The Pearson correlation coefficients are presented in the upper triangle, while in the lower triangle the correlation coefficients are colour-coded.}
\label{fig:corr_matrix_physical_pars}
\end{figure}

\begin{figure*}
\centering
\includegraphics[width=1.0\textwidth]{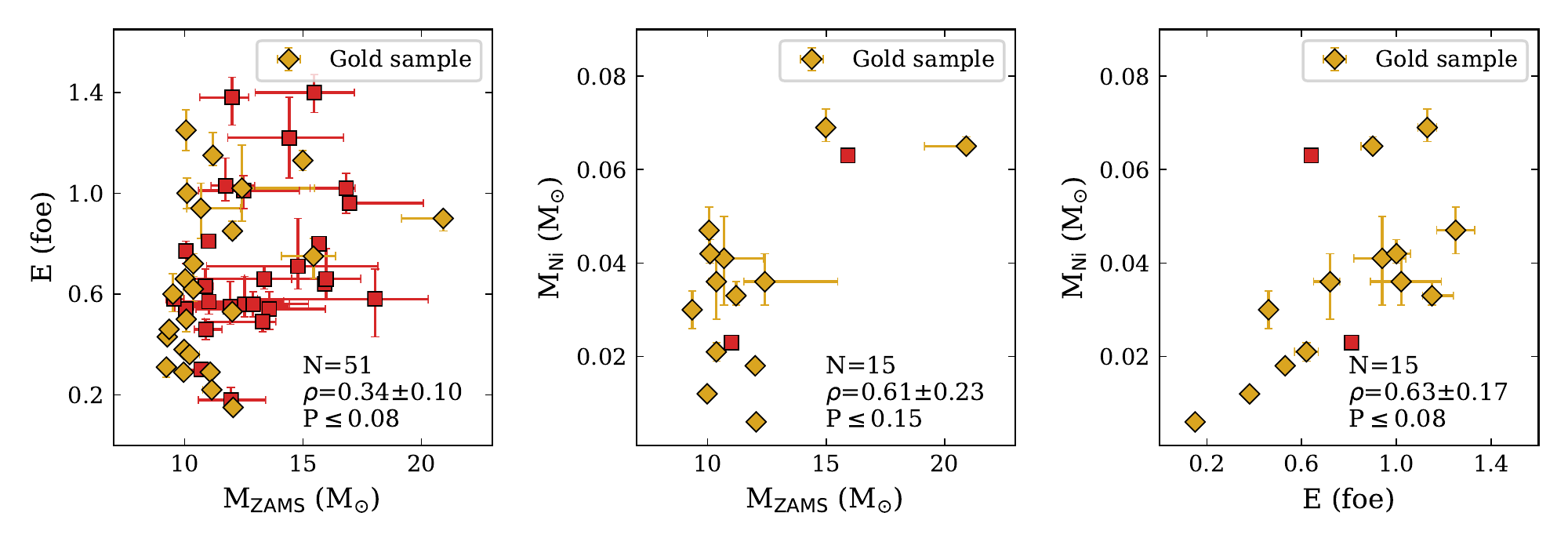}
\caption{Correlations between three physical parameters: \mzams, \e, and \mni. Each subplot contains the number of events ($N$), the Pearson correlation coefficient (\p), and the probability of detecting a correlation by chance ($P$) using the full sample. Yellow markers indicate gold events.}
\label{fig:corr_physical_pars}
\end{figure*}

Figure~\ref{fig:corr_matrix_physical_pars} shows the correlation matrix of the physical parameters. Given that the progenitor models used were calculated assuming single-star evolution with a standard wind efficiency, all the progenitor parameters (\mzams, \mej, \mhenv, and \ra) are strongly correlated.
In this context, \mzams\ is the only independent variable related to a unique pre-SN structure. This structure is characterised by \mej, \mhenv, and \ra\ which increase almost monotonically with \mzams\ within the range of initial masses of our pre-SN models \citepalias[see][their Fig.~2]{martinez+20}. 

In Fig.~\ref{fig:corr_physical_pars} (left panel), \mzams\ is
plotted against \e. It is found that these physical parameters show a weak correlation (\p\,=\,0.34\,$\pm$\,0.10, $N$\,=\,51) in the sense that higher \mzams\ progenitors display higher energy explosions. 
Most studies compare \e\ with \mej\ instead of \mzams. We also find a weak correlation between \mej\ and \e\ (\p\,=\,0.36\,$\pm$\,0.11) which was expected because of the tight relation between \mzams\ and \mej.

Similar trends were also inferred by previous studies that obtained physical parameters from observations via hydrodynamical modelling or analytic scaling relations \citep[e.g.][]{hamuy03,morozova+18,utrobin+19}, although most of these studies find stronger correlations. 
This difference could be associated to the type of pre-SN structure adopted to initialise the explosion model. 
Particularly, works assuming polytropic progenitor models find tight relations between \mej\ and \e, while weaker trends are found using pre-SN models from standard single-star evolution calculations \citep[see Fig.~6 by][for a comparison of different studies]{burrows+21}.
This may be due to the contribution of two factors. On the one hand, polytropic models allow the construction of progenitor structures over a wide range of pre-SN masses, hydrogen-rich envelope masses, and progenitor radii. On the other hand, parameter degeneracy ---where different physical parameters sometimes produce similar LCs and spectral properties during the recombination phase--- sometimes admits solutions for larger masses and higher explosion energies simultaneously, given that the effect of both parameters on LCs and photospheric velocities is cancelled.
However, the small range of masses predicted by standard single-star evolution, in addition to the wide range of inferred explosion energies, leads to scatter in the mass--energy relation in the current work.
While the dispersion of this relation is large, our results show a lack of high-mass progenitors. This may imply an important constraint to the explosive end of more massive stars.
Most of the inferred progenitor masses range between 9 and 13~\ms\ and are consistent with models of varied energy covering the range of 0.15 to 1.40~foe. That is, while the inferred explosion energies span the full range explored in our models, the constrained masses generally only sample the low-mass end.

The middle panel of Fig.~\ref{fig:corr_physical_pars} indicates a positive trend between \mzams\ and \mni\ similar to previous findings in the literature \citep[e.g.][]{hamuy03,utrobin+19}. We find a correlation coefficient of \p\,=\,0.61\,$\pm$\,0.23; however, given the low number of events ($N$\,=\,15), this trend may be driven by the three objects with high \mzams\ and \mni.
We also note that low-\mzams\ progenitors cover a large range of \mni\ values between $\sim$0.005 and 0.05~\ms. These objects represent 12 of the 15~\sneii\ with derived \mni. The other three progenitors are compatible with stars more massive than 15~\ms, and all have \mni\ values above 0.06~\ms.
That is, modelling of our \snii\ observations shows that progenitors with \mzams\ larger than 15~\ms\ exclusively show high \Ni\ masses.

In the right panel of Fig.~\ref{fig:corr_physical_pars}, the explosion energy is plotted against \mni. It is found that these parameters show a positive correlation (\p\,=\,0.63\,$\pm$\,0.17, $N$\,=\,15) with higher energy explosions showing more \Ni. 
\Ni\ originates from explosive nucleosynthesis during a SN explosion. After this explosive burning phase, the innermost layers of the ejecta can become bound and fall back onto the compact remnant carrying a significant fraction of \Ni.
Thus, in principle, the amount of \Ni\ powering the radioactive tail phase is not that produced during explosive nucleosynthesis, which may produce a bias in the \e$-$\mni\ correlations.
However, fallback is expected to be less important at higher explosion energies, and in our case, only a small number of \sneii\ are compatible with low explosion energies. Therefore, we do not expect a large bias in our trend. In Sect.~\ref{sec:energy} we discuss the physical origin of this correlation and its connection with the observed diversity of \sneii.

%------------------------------------------------------
\subsection{Unveiling the \snii\ physical parameter space of observed explosions}

Throughout the previous subsections we analyse several correlations between physical and observed properties of \sneii. Here we focus on examining the parts of the parameter space that are not reached with our observations because this may afford significant constraints to the evolution and explosion of massive stars. The large number of \sneii\ in the CSP-I sample allows such an analysis.

Observed \optd\ ranges from 42 and 146~days \citepalias{martinez+21a}.
While the \optd\ values in our sample longer than 100~days are well reproduced by models for a variety of \mhenv\ (from the shortest to the largest \mhenv\ within our grid of models; see Fig.~\ref{fig:corr_optd_std}), the \sneii\ in the full sample displaying the shortest \optd\ ($\sim$80~days) are always found for low \mhenv, specifically \mhenv\,$\lesssim$\,8~\ms. Therefore, such a short \optd\ is not found for \mhenv\,$\gtrsim$\,8~\ms.
Short \optd\ for high \mhenv\ could be obtained for sufficiently energetic explosions, but this does not seem to be the case here.
As discussed previously, we also note that there are models with \optd\ values much larger than any of those found for  observed SN~II to date.

\mbolend\ and \e\ show a tight correlation with a low dispersion, and therefore less luminous \sneii\ are not found for higher explosion energies and vice versa (Fig.~\ref{fig:corr_mbol_std}).
Additionally, \sneii\ with \mbolend\ dimmer than around $-$16~mag are not consistent with \Ni\ masses in the ejecta higher than $\sim$0.05~\ms.

We already mentioned the strong dependency of \mboltail\ on \mni; therefore, more (less) luminous \sneii\ during the radioactive tail phase are not compatible with low (high) \Ni\ masses.
Moreover, some regions of the \mboltail-\e\ parameter space are not covered, specifically those for bright (dim) radioactive tails and low (high) explosion energies. However, this absence is related to the physical connection between \e\ and \mni\ (Sect.~\ref{sec:corr_physical_pars}). Therefore, higher energy explosions with low \Ni\ masses (and vice versa) are also missing.
Furthermore, our results show a lack of high-mass progenitors particularly with low explosion energies, similar to previous studies \citep[see][and references therein]{morozova+18}.

Lastly, we analysed the parameter space that is not covered by the observed \stwo\ declination rate. Although some dispersion is found, it is seen that the largest (smallest) values of \stwo\ are not compatible with low (high) explosion energies (see Sect.~\ref{sec:discussion} for explanation). At the same time, while only one slowly declining \snii\ with an inner distribution of \Ni\ within the ejecta is found, more steeply declining \sneii\ are not consistent with \Ni\ extended to the outer regions of the ejecta.

%------------------------------------------------------
\section{Results from non-standard pre-SN evolution}
\label{sec:nonstd}

The grid of explosion models used in \citetalias{martinez+21b} employs pre-SN structures from stellar evolution calculations to initialise the explosion, which were calculated using the standard prescription for the wind mass loss defined in the \texttt{MESA} code (see Sect.~\ref{sec:models}).
While the CSP-I \snii\ sample consists of 74 objects, in \citetalias{martinez+21b} we derived physical and explosion properties for 53 events. Some \sneii\ were excluded for the various reasons outlined in \citetalias{martinez+21b}, and these are not repeated in the current work.
Here we only mention the two relevant (for the current analysis) \sneii\ that were excluded: SN~2006Y and SN~2008bu.
These SNe present atypically short \optd\ values of 64\,$\pm$\,4~days and 52\,$\pm$\,7~days for SN~2006Y and SN~2008bu, respectively \citepalias{martinez+21a}. Given that none of our LC models present such short \optd\ values, SN~2006Y and SN~2008bu could not be well fitted. 

Table~\ref{table:kda} shows that the explosion energy is the parameter that produces the largest impact on \optd. A higher explosion energy would lead to a shorter \optd, that is, in the direction needed to model the previous two short-plateau \sneii. However, the more energetic explosions also produce more luminous \sneii\ that display higher velocities, differing even further from observations.
Therefore, here we present new pre-SN models evolved with a higher mass-loss rate, which reduces the extent of the hydrogen-rich envelope at the time of core collapse, and therefore the \optd. With this assumption, we are able to model these two short-plateau \sneii\ \citep[see also][]{hiramatsu+21}.
Furthermore, in order to analyse the extent of variation in the correlations when the pre-SN models change, we modelled all the well-observed \sneii\ in the CSP-I sample (previously modelled in \citetalias{martinez+21b} using standard stellar models) employing the explosion models with additional mass stripping.

%------------------------------------------------------
\subsection{New progenitor and explosion models}
\label{sec:nonstd_models}

\input{presn_models.tab}

In order to assess the physical parameters for the short-plateau \sneii\ in the CSP-I sample, we constructed new non-rotating solar-metallicity pre-SN RSG models using the stellar evolution code \texttt{MESA} version 10398 \citep{paxton+11,paxton+13,paxton+15,paxton+18,paxton+19}.
Given that we want to reproduce the short-plateau phase of SN~2006Y and SN~2008bu, higher wind efficiencies during the evolution of the progenitor star were assumed to reduce the final mass of the hydrogen-rich envelope\footnote{However, it is important to note that this stellar wind efficiency was used to mimic any mechanism that may produce additional envelope stripping (including for instance binary interactions and eruptive mass loss).}.
Stars were evolved for three \mzams\ values (10, 12, and 14~\ms) to obtain models with different pre-SN radii. The final radii range from 455 to 851~\rs, which is a good starting point for our following analysis.
We used the `Dutch' recipe for mass loss defined in \texttt{MESA}. The wind scaling factor ($\eta$) linearly modifies the mass-loss rate and is equal to unity in the standard pre-SN models. Here, $\eta$ is arbitrarily chosen to produce \mhenv\ in the range of $\sim$3.5$-$7.1~\ms, the latter value being the lowest \mhenv\ in the standard pre-SN models.
The other stellar evolution parameters took the same values as summarised in Sect.~\ref{sec:models}.
Table~\ref{table:presn_models} lists the physical characteristics of the pre-SN models with enhanced mass loss, together with the standard pre-SN models.
We then computed a grid of synthetic bolometric LCs and photospheric velocities using a 1D hydrodynamical code \citep{bersten+11} in the same manner as we constructed our standard set of models used in \citetalias{martinez+21b}.
We varied the explosion energy between 0.2 and 1.5~foe and \mni\ between 0.005 and 0.08~\ms\ for each progenitor model. In addition, three values of \mix\ for each explosion model were considered to account for the effect of the spatial distribution of \Ni\ within the ejecta: out to the 20\%, 50\%, and 80\% of the pre-SN structure in mass coordinates.

We used the same fitting procedure as in \citetalias{martinez+21b}, which is based on MCMC methods and allows one to find the posterior probability of the model parameters given the observations (see Sect.~\ref{sec:models}). 
In \citetalias{martinez+21b}, we used six parameters to model the observables: the explosion epoch (\texp), scale, \mzams, \e, \mni, and \mix.
The pre-SN models used in \citetalias{martinez+21b} were calculated assuming the standard prescriptions for single-star evolution; consequently, \mzams\ is the only independent variable related to a unique pre-SN structure.
Here, we obtained different pre-SN structures for the same \mzams\ given that we also varied the efficiency of wind mass loss. Therefore, we included $\eta$ as an additional parameter to break the previous degeneracy between \mzams\ and the pre-SN structures.
We used uniform priors for six parameters: \texp, \mzams, $\eta$, \e, \mni, and \mix. The sampler is allowed to run within the observational uncertainty of the \texp\ (see Table~1 from \citetalias{martinez+21a}) and within the parameter space for the other parameters.
For the scale parameter, we used a Gaussian prior centred at one, with a standard deviation controlled by the uncertainty in the distance and host-galaxy extinction (see \citetalias{martinez+21b} for details).
We employed 400 walkers and 10000 steps per sampler, with a burn-in period of 5000 steps. The walkers were randomly initialised, covering the entire parameter space.

%------------------------------------------------------
\subsection{Correlations between physical and observed parameters using non-standard pre-SN models}
\label{sec:corr_nonstd}

\input{corr_comparison.tab}

\begin{figure*}
\centering
\includegraphics[width=0.97\textwidth]{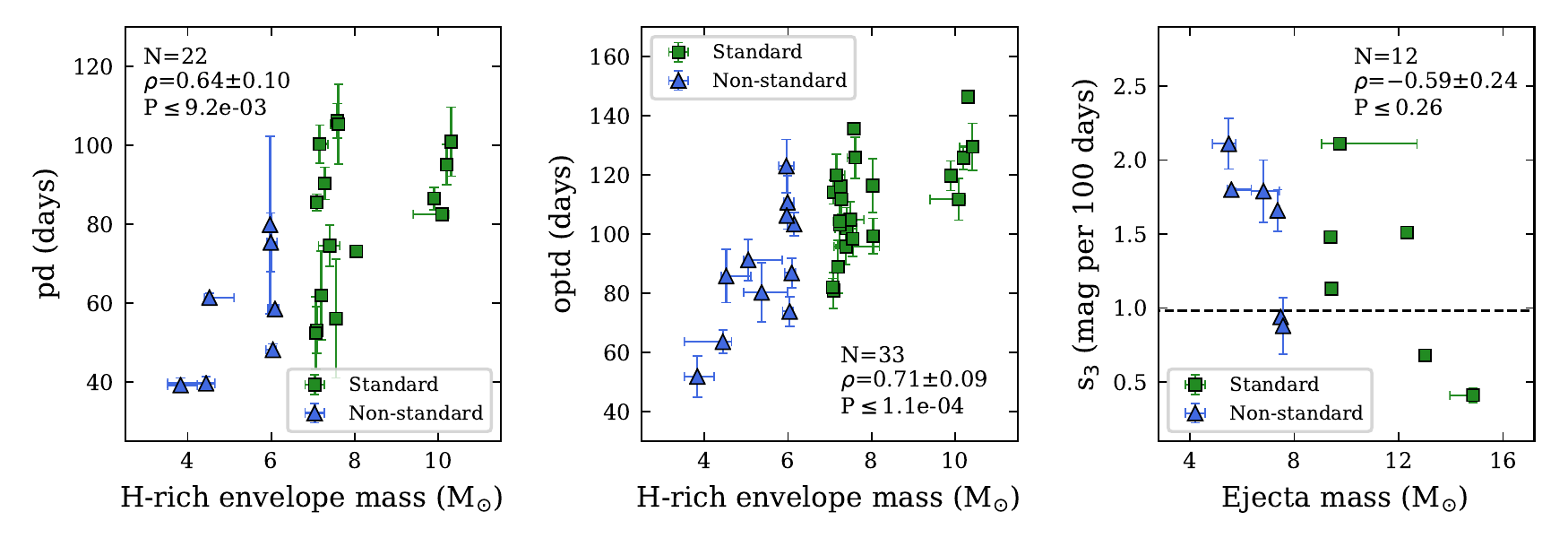}
\caption{Correlations between \mhenv\ and \pd\ (left panel), \mhenv\ and \optd\ (middle panel), and \mej\ and \sthr\ (right panel) using the results from non-standard pre-SN models. In the right panel, the dashed line indicates the expected declination rate for full trapping of $^{56}$Co decay. Each subplot contains the number of events ($N$), the Pearson correlation coefficient (\p), and the probability of detecting a correlation by chance ($P$). Blue triangles show the results of using non-standard pre-SN models, while green squares indicate results from standard evolution. SNe labelled as `non-standard' have Bayes factors greater than 10$^{1/2}$.}
\label{fig:corr_hmass_nonstd}
\end{figure*}

\begin{figure}
\centering
\includegraphics[width=0.50\textwidth]{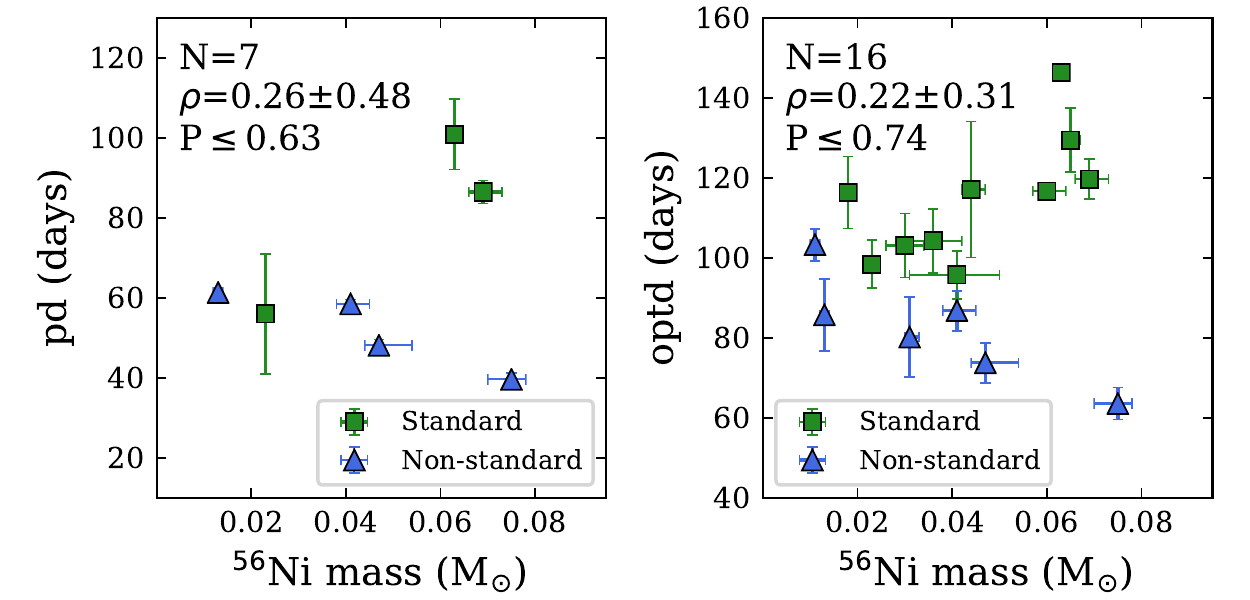}
\caption{Correlations between physical and observed parameters using the results from non-standard pre-SN models. \textit{Left panel:} \mni\ versus \pd. \textit{Right panel:} \mni\ versus \optd. Each subplot contains the number of events ($N$), the Pearson correlation coefficient (\p), and the probability of detecting a correlation by chance ($P$). Blue triangles show the results obtained using non-standard pre-SN models, while green squares indicate results from standard evolution. SNe labelled as `non-standard' have Bayes factors greater than 10$^{1/2}$.}
\label{fig:corr_ni_mix_nonstd}
\end{figure}

The bolometric LC and expansion velocity models presented in the previous section were designed to determine the physical properties of SN~2006Y and SN~2008bu, that is, the two shortest-plateau \sneii\ in the CSP-I sample.
The posterior distributions of the parameters were constructed using the MCMC fitting procedure described in Sect.~\ref{sec:nonstd_models}. In Appendix~\ref{app:non_std}, models drawn from the posterior distribution in comparison with the observations are presented for these two \sneii.

The short-plateau LC and expansion velocities of SN~2006Y are well reproduced by models with \mej\,$\simeq$\,5.5~\ms, \mhenv\,$\simeq$\,4.4~\ms, \ra~$\simeq$~460~\rs, and \e\,$\simeq$\,1.2~foe.
The progenitor structure is the result of the evolution of a star with an initial mass of 10.3~\ms\  and a high wind scaling factor of $\eta$\,$\sim$\,8.
The LC of SN~2006Y was recently modelled by \citet{hiramatsu+21}. We find a progenitor radius consistent with the results of these latter authors, although they found a less massive hydrogen-rich envelope mass of 1.7~\ms\ and a lower explosion energy of 0.8~foe. 
These differences may be attributed to the wider range of progenitor properties studied by \citet{hiramatsu+21} ---particularly \mzams\ and the wind scaling factor--- or to the fact that they determine progenitor and explosion parameters without using any spectral information in their modelling \citepalias{martinez+20}.
The observations of SN~2008bu are consistent with \mej\,$\simeq$\,5.2\,\ms, \mhenv\,$\simeq$\,3.8\,\ms, \ra\,$\simeq$\,630\,\rs, and \e\,$\simeq$\,0.5\,foe.
The pre-SN model used to reproduce the observations of SN~2008bu is the consequence of the evolution of a star with \mzams\,$\simeq$\,11.5~\ms, with a wind scaling factor of around 6.
Therefore, as expected, SN~2006Y and SN~2008bu are reasonably well reproduced with models calculated with significant envelope stripping.
These results are reported in Appendix~\ref{app:non_std}. They are characterised by the median of the marginal distributions, while we adopt the 16th and 84th percentiles as our lower and upper uncertainties, respectively. We also include estimates of \mej, \mhenv, \mh, and \ra. These values are not model parameters, and are therefore not fitted. These values are interpolated from the \mzams\ and $\eta$ derived in the fitting.

Initially, the pre-SN models with additional mass stripping together with their corresponding explosion models (Sect.~\ref{sec:nonstd_models}) were calculated to reproduce the observations of SN~2006Y and SN~2008bu because of their short optically thick phases.
However, we used these synthetic bolometric LCs and expansion velocities to model several other \sneii\ in the CSP-I sample that were already modelled in \citetalias{martinez+21b} in order to determine possible variations in the correlations found in Sect.~\ref{sec:corr_std}.
Given the degeneracies in inferring physical properties from LC modelling, we were able to obtain more than one set of physical parameters for the same SN.

We explored new solutions for those \sneii\ in the CSP-I sample that fulfil the following criteria: (a) enough photometric observations to cover the photospheric phase and at least the beginning of the transition to the radioactive tail; and (b) at least one measurement of the \ion{Fe}{ii} velocity. This information is crucial to determining the hydrogen-rich envelope mass (among other physical properties).
We find 31~\sneii\ (without considering SN~2006Y and SN~2008bu) that fulfilled our criteria. For these 31~\sneii, we used the MCMC procedure described in Sect.~\ref{sec:nonstd_models} to determine a set of physical parameters consistent with partially stripped progenitors.
For each SN, we then compared their observations with models drawn from the posterior distribution. For some SNe, we find large differences between models and observations, specifically in the duration of the optically thick phase.
However, for some others, the models with additional mass stripping are able to reproduce the observations.
We quantified the significance for this solution over that from \citetalias{martinez+21b} using the Bayes factor ($B$). According to \citet{jeffreys1998}, $B$~>~10$^{1/2}$ implies that the support for the solution with reduced hydrogen-rich envelope is substantial, strong if $B$ is between 10 and 10$^{2}$, and decisive if $B$~>~10$^{2}$.
Following the above statistical analysis, we find that the solution with additional mass lost is favoured against the standard solution ($B$~>~10$^{1/2}$) for nine \sneii. Together with SN~2006Y and SN~2008bu, we therefore reach a sample of 11~\sneii\ that are better fit with these higher mass-loss models.
Their physical parameters, in addition to comparisons between models and observations, are presented in Appendix~\ref{app:non_std}. 

These results were then used to reanalyse the correlations between observed and physical parameters. The results for the 11~\sneii\ that favour non-standard pre-SN models were used. At the same time, the results from standard single-star evolution (i.e. from \citetalias{martinez+21b}) were employed for the remaining \sneii.
In general, similar trends to those presented in Sect.~\ref{sec:corr_std} were found. The largest differences are described below.

Figure~\ref{fig:corr_hmass_nonstd} presents \mhenv\ against \pd\ (left panel), \mhenv\ against \optd\ (middle panel), and \mej\ against \sthr\ (right panel).  \mhenv\ shows strong correlations with \pd\ and \optd. The correlation coefficients are \p\,=\,0.64\,$\pm$\,0.10 ($N$\,=\,22) and \p\,=\,0.71\,$\pm$\,0.09 ($N$\,=\,33), respectively. In addition, a moderate trend is found between \mej\ and \sthr\ with \p\,=\,$-$0.59\,$\pm$\,0.24 ($N$\,=\,12).
In Table~\ref{table:corr_comparison}, the strengths of correlations are compared between standard and non-standard results for those correlations that show the largest differences.
We find that for these three cases, the strengths of the correlations are now significantly greater than in Sect.~\ref{sec:corr_std}, where we analysed the results from standard evolutionary models.
Figure~\ref{fig:corr_ni_mix_nonstd} shows \mni\ against \pd\ and \optd. 
In Sect.~\ref{sec:results}, we show a trend between these parameters in the sense that \sneii\ with higher \Ni\ masses develop a longer plateau and optically thick phases.
However, we now find weak trends (see Table~\ref{table:corr_comparison}). We note at least one obvious outlier in these plots (SN~2006Y), with the correlations being stronger when this event is removed. SN~2006Y presents the shortest \pd\ and \optd\ with the largest \mni\ estimate (\mni\,=\,0.075).
Therefore, this shows that a large amount of \Ni\ present in the SN ejecta does not necessarily imply long \pd\ and \optd.

The rest of the correlations show little or no variation between standard and non-standard evolution models and they are not discussed further.
The differences found in the correlations show the impact of the parameter degeneracy on the results, given that this may alter the strength of some correlations.
However, in the cases where the parameter degeneracy admitted different results, only the strength of correlations changed (Table~\ref{table:corr_comparison}).
The parameter degeneracy did not produce new trends, nor did it cause the disappearance of previous correlations.
We note that the trends between \mni\ and \pd\ and \mni\ and \optd\ disappear because of one outlier (SN~2006Y; see Fig.~\ref{fig:corr_ni_mix_nonstd}) and not because of parameter degeneracies, given that SN~2006Y was modelled only once (SN~2006Y could not be modelled with our grid of standard models).
The correlation matrix between observed and physical parameters using non-standard pre-SN models is presented in Appendix~\ref{app:non_std}.

We determined the effect of each physical parameter on \snii\ observables using the models constructed with the moderately stripped progenitors presented in Sect.~\ref{sec:nonstd_models}. We measured observables from the non-standard LC and photospheric velocity models as described in Sect.~\ref{sec:models}. We then performed a KDA in a similar manner to in Sect.~\ref{sec:results}.
For the KDA, we used the standard models and those with additional mass stripping together. The inclusion of models from non-standard pre-SN calculations raises a problem, given that there is more than one model for several sets of physical parameters.
This issue is solved by including the wind efficiency $\eta$ as an additional physical parameter for the KDA. Values of $\eta$ equal unity for the standard models, and its corresponding value (see Table~\ref{table:presn_models}) for each model with additional mass stripping.
\mzams\ and $\eta$ are highly related to the ejecta mass. As a consequence, the combination of \mzams\ and $\eta$ represent, at first order, the effect of the ejecta mass.
Two major changes are found with respect to the results presented in Table~\ref{table:kda}. We find a lower influence of the explosion energy on \optd, while both \mzams\ and $\eta$ together present a higher effect than that corresponding to \mzams\ in Table~\ref{table:kda}.
Moreover, the impact of \mzams\ and $\eta$ on \sthr\ increases considerably, reaching a relative importance of 0.73 against the estimate of 0.58 found for \mzams\ in Table~\ref{table:kda}.

%------------------------------------------------------
\section{Discussion}
\label{sec:discussion}

%------------------------------------------------------
\subsection{The effect of explosion energy}
\label{sec:energy}

Following the analysis in Sect.~\ref{sec:results}, it seems that the majority of the diversity in \snii\ LCs can be confidently ascribed to differences in explosion energy when standard single-star evolution is assumed.
A higher explosion energy leads to more luminous \sneii\ with higher expansion velocities, which cool and recombine the ejecta more rapidly, producing a shorter plateau and optically thick phases.
Moreover, faster expansion of the SN ejecta induces lower densities at earlier epochs. At sufficiently low ejecta densities, the cooling wave ---which is responsible for the plateau--- cannot be formed, resulting in steeper plateau phases \citep{grassberg+71,blinnikov+93}.
A faster expansion also produces a more rapid cooling of the SN ejecta, which is consistent with the picture of faster declining \sneii\ having redder colours at late times \citep{dejaeger+18}. 
Additionally, higher energy explosions produce more \Ni, which cause increased variation in the late-plateau and tail brightness (see discussion below). We now compare our results with those found in previous studies.

The declination rate during the plateau (\stwo) is one of the parameters that show great diversity and a continuum of values \citep[\citetalias{anderson+14_lc};][]{sanders+15,galbany+16,valenti+16,rubin+16,dejaeger+19}. 
Theoretical studies suggest that fast declining \sneii\ are produced by reducing the progenitor hydrogen-rich envelope mass \citep[e.g.][]{blinnikov+93,hillier+19,pessi+19}. 
Observational studies support this scenario based on correlations between \stwo\ and observed parameters that are mostly related to the progenitor envelope mass, such as the plateau and optically thick phase durations \citepalias{anderson+14_lc,gutierrez+17II}.
While low hydrogen-rich envelope masses satisfactorily produce the low ejecta densities necessary to avoid the cooling wave, these low densities can also be achieved by sufficiently high explosion energies.
Our results show a lack of correlation between \mhenv\ and \stwo\ (see Fig.~\ref{fig:corr_hmass_s2_std}), but this analysis does not contradict the prediction that small hydrogen-rich envelope masses produce fast-declining LCs; indeed some of the fastest decliners are consistent with the lowest \mhenv\ values of our models. Rather, this analysis may imply that \mhenv\ is not the main driver of the diversity observed in plateau declination rates.
However, our grid of pre-SN models was calculated assuming standard prescriptions for the wind-mass loss, which does not account for highly stripped progenitors. While this may bias our conclusions, the results using the explosion models from moderately stripped progenitors (non-standard models) also suggest a lack of correlation between \mhenv\ and \stwo. We note that only a narrow range of \mhenv\ was studied here and a more exhaustive analysis is needed.

It has also been suggested that \Ni\ mass can explain the \stwo\ diversity. Figure~\ref{fig:mbolend} (middle panel) indicates that the late-plateau luminosity increases if \Ni\ is present in the SN ejecta, yielding a shallower slope. 
In addition, the left panel of Fig.~\ref{fig:s2} clearly shows the \Ni\ effects on \stwo, which can vary by up to $\sim$1~mag per 100 days for the models presented in that plot.
This was also proposed in theoretical studies by \citet{bersten13phd} and \citet{kozyreva+19}.
From an observational point of view, \citet{nakar+16} analysed a sample of 24~\sneii\ and concluded that \Ni\ can flatten the plateau phase. 
While low-\mni\ events can produce fast-declining \sneii, we find no clear relation between \mni\ and \stwo\ for the CSP-I \snii\ sample, in accordance with the results from \citetalias{gutierrez+17II}.

An alternative scenario for the diversity of \stwo\ includes the interaction of the SN ejecta with a dense CSM surrounding the progenitor star.
The presence of CSM can boost the early-time luminosity, thus resulting in rapidly declining LCs.
\citet{morozova+17} show that the multi-band LCs of slow- and fast-declining \sneii\ are well reproduced by RSG explosions that collide with a dense CSM of different properties \citep[see also][]{hillier+19}.
However, for low to moderate CSM masses, only the early LC is affected.
Without considering the effects of CSM interaction, we find that our models reproduce most well-sampled \sneii\ after 30~days post-explosion, including fast declining \sneii. Therefore, we do not expect the effects of CSM interaction to dominate the late-time declination rate.

\citet{kozyreva+19} suggest that more energetic events evolve more rapidly and have faster declination rates if the contribution of \Ni\ is ignored. This is somewhat similar to our findings.
In Sect.~\ref{sec:corr_std} we present our findings of evident correlations between \e\ and \stwo, which imply that fast-declining \sneii\ in the CSP-I sample are consistent with high-energy explosions.
Additionally, from the analysis of the model parameters through the KDA (Table~\ref{table:kda}), we find that the explosion energy is the physical parameter that most influences the \stwo\ declination rate.
For these reasons, we conclude that the explosion energy is the main driver of \stwo\ diversity, that is, under the assumption of standard single-star evolution. 

Previous theoretical studies found that the explosion energy is strongly related to \pd\ and \optd, because more energetic explosions lead to higher expansion velocities, and therefore more rapid cooling and hydrogen recombination \citep[e.g.][]{bersten13phd,dessart+13}.
This is supported by our results given that we find moderate anti-correlations between \e\ and \pd\ and also between \e\ and \optd. In addition, previous observational studies found an anti-correlation between \pd\ and expansion velocities, which are closely related to the explosion energy \citep[][\citetalias{gutierrez+17II}]{faran+14a}.
The \pd\ is also affected by the mass of the hydrogen envelope as is further discussed in Sect.~\ref{sec:hmass}.

Very strong correlations between the explosion energy and the \snii\ brightness at the photospheric phase are found, in line with theoretical predictions \citep{bersten13phd}. This is also seen in observational studies, where it is found that more luminous \sneii\ develop higher expansion velocities, and must therefore arise from more energetic explosions \citep[e.g.][\citetalias{gutierrez+17II}]{hamuy+02}.
There are exceptions to this trend, with \sneii\ displaying low velocities and high luminosities \citep{rodriguez+20}, although these observational features are probably due to the interaction of the SN ejecta with a massive CSM.
In the current paper, we also find a very strong correlation between \e\ and \mboltail\ \citepalias[see also][]{gutierrez+17II}.
However, \mboltail\ strongly depends on the amount of \Ni\ present in the ejecta \citep{woosley+89} and not ---directly--- on the explosion energy. Moreover, we find that \mboltail\ is not greatly impacted by explosion energy (Fig.~\ref{fig:mboltail}). Therefore, we should not expect a correlation between \mboltail\ and \e.
The caveat here is that we treated \e\ and \mni\ as purely independent variables while it appears that they may be closely connected.

\Ni\ production takes place during a SN explosion. After core collapse, a shock wave forms accelerating and heating the stellar material above the nascent proto-neutron star. 
The total amount of \Ni\ synthesised depends on the mass of the ejecta exposed to temperatures above 5\,$\times$\,10$^{9}$~K. Above this temperature, explosive Si burning occurs where \Ni\ dominates the production of nuclear species \citep{woosley+95,thielemann+96,umeda+02}.
For a low explosion energy, the temperature needed for explosive Si burning is only reached in the innermost layers of the SN ejecta, while a higher explosion energy increases the mass exposed to high temperatures \citep{sukhbold+16}. Therefore, a positive correlation between \e\ and \mni\ is expected. 
Our findings show a strong correlation between these two physical parameters (see Fig.~\ref{fig:corr_physical_pars}).
In addition, a correlation between \mni\ and \ion{Fe}{ii} $\lambda$5169 velocities is found (\p\,=\,0.58\,$\pm$\,0.21), in accordance with previous observational studies \citep[e.g.][\citetalias{gutierrez+17II}]{hamuy03,spiro+14}.
The correlations between \e, \mni, \mbolend, and expansion velocities support the hypothesis that the explosion energy is the driver of these relations: more energetic explosions produce brighter \sneii\ with faster expansion velocities and more \Ni, which increments the tail luminosity.

\citetalias{gutierrez+17II} found an  inverse correlation between expansion velocities and the strength of metal lines and concluded that higher energy explosions produce higher temperatures for a longer time, leading to lower metal-line pEWs. In the current analysis, we recover those trends using the pEWs of \ion{Fe}{ii} $\lambda$4924 and \ion{Fe}{ii} $\lambda$5018. 
The emission component of \ha\ and the absorption component of \hb\ show the opposite behaviour, namely moderate correlations with \e. This was also inferred by \citetalias{gutierrez+17II}.

Our study suggests that the explosion energy is the key parameter underlying \snii\ diversity, which raises the question of what determines the energy of the explosion.
Massive stars develop an iron core in the centre during the last stage of nuclear burning. At the end of massive-star evolution, the iron core becomes gravitationally unstable and starts to collapse. The collapse of the central parts is halted when nuclear densities are reached, at which point this collapse rebounds, producing a shock wave. However, this shock wave is not sufficiently energetic and fails to trigger the explosion \citep[see][for a review]{janka12}. 
The mechanism of the energy deposition into the envelope to reverse the stalled shock into explosion is still debated, but the so-called `delayed neutrino-driven mechanism' is the most favoured scenario \citep{bethe+85,janka12}.
The gravitational binding energy of the neutron star ($\gtrsim$10$^{53}$~erg) is released as neutrinos when the core of a massive star collapses to a neutron star. The interaction of a fraction of the neutrino flux with the material of the mantle powers the explosion \citep[for a recent review]{colgate+66,janka+16}.
The rapid rotation of the collapsing core and magnetic fields may also play a role \citep{leblanc+70,bisnovatyi-kogan71}.

Many observational and theoretical works have analysed a possible connection between \e\ and \mzams\ (or \mej).
From the observational point of view, \citet{hamuy03} and \citet{pejcha+15a} used the analytic scaling relations of \citet{litvinova+85} and found a positive correlation between \e\ and \mej.
Hydrodynamical modelling of several \sneii\ reveals a general trend with more massive progenitors undergoing more energetic explosions, although with large scatter \citep[e.g.][]{pumo+17,morozova+18,eldridge+19b,utrobin+19,ricks+19,martinez+19}. The same is found in the current analysis (Sect.~\ref{sec:corr_physical_pars}).
However, explosion models based on the neutrino mechanism have found little evidence of a correlation between \e\ and the progenitor mass \citep[e.g.][]{ugliano+12,pejcha+15,ertl+16,sukhbold+16}. None of these studies recover a monotonic increase in \e\ with \mzams.
Recently, 3D core-collapse simulations found that lower mass progenitors produce lower explosion energies, while higher mass progenitors experience higher energy explosions \citep{burrows+20}.
\citet{burrows+21} show that the general trend of the mass--energy relation inferred by observational studies is reproduced by theoretical models, although some scatter is present both in observations and models.
In addition, \citet{burrows+21} argue that a diversity of explosion energies, among other physical properties, could be obtained for the same initial pre-SN structure.

%------------------------------------------------------
\subsection{The effect of hydrogen-rich envelope mass}
\label{sec:hmass}

During the plateau phase, hydrogen recombination occurs at different layers of the ejecta as a recombination wave recedes (in mass coordinate) through the expanding envelope \citep{grassberg+71,bersten+11}. Because the opacity is dominated by electron scattering, it decreases outwards from the recombination front allowing the radiation to escape. 
Therefore, the duration of the plateau phase is connected to the hydrogen-rich envelope mass at the time of core collapse, among other physical properties.
Short plateau \sneii\ are usually associated with low hydrogen envelope masses (see Sect.~\ref{sec:corr_nonstd}), while the opposite is found for long-plateau \sneii\ \citep[e.g.][]{litvinova+83}.
Our results are consistent with this picture given that we find moderate trends between \mhenv\ and both \pd\ and \optd.

\citetalias{anderson+14_lc} and \citetalias{gutierrez+17II} introduced an additional parameter as a tracer of the hydrogen-rich envelope mass. Both studies found that the declination rate during the radioactive tail phase (\sthr) is usually higher than that expected from the decay rate of $^{56}$Co if full trapping of gamma-ray photons is assumed \citep{woosley+89}. 
In addition, \citetalias{anderson+14_lc} and \citetalias{gutierrez+17II} found that \sthr\ is strongly related to $OPTd$ and $Pd$ (the uppercase acronyms refer to the definitions from \citetalias{anderson+14_lc} and \citetalias{gutierrez+17II} which are slightly different from ours), in the sense that shorter plateau \sneii\ display higher declination rates in the radioactive phase.
This may suggest that, in these cases, \mej\ is too small for full trapping of the gamma-rays, which causes higher \sthr\ values. 
While in principle we recover a weak trend between \mhenv\ and \sthr, the significance of the correlation is low (see Fig.~\ref{fig:corr_s3_std}, right panel).

The previously mentioned correlations are much stronger when using the results from moderately stripped progenitors (Sect.~\ref{sec:corr_nonstd}), which leads to the following conclusions: 
(1) \sthr\ is affected by the ejecta mass, which is consistent with the analysis presented in Table~\ref{table:kda} and Sect.~\ref{sec:corr_nonstd}, where we find that the mass is the physical parameter causing the greatest effect on \sthr, and with previous studies in the literature; (2) additional mass stripping than predicted by standard single-star evolution is necessary to reproduce the great dependency of the hydrogen-rich envelope mass with \pd\ and \optd\ (see Fig.~\ref{fig:corr_hmass_nonstd}). 
Moreover, in \citetalias{martinez+21b}, we found a clear inconsistency between our \mzams\ distribution and a Salpeter IMF, where we invoked additional mass loss as one possible explanation (see \citetalias{martinez+21b} for details).
Therefore, our findings seem to suggest that higher mass-loss rates are required to reproduce SN~II observations.
However, recent studies suggest that prescriptions for wind mass-loss rates used in stellar evolution already overestimate the mass loss by winds \citep{puls+08,smith14,beasor+20}.
The increment of the wind mass-loss rate in our models was used to mimic any mechanism producing additional envelope stripping. Alternatively, stellar eruptions, rotation, and/or mass transfer by Roche-Lobe overflow in a binary system should be studied more thoroughly.

The hydrogen-rich envelope mass is an important parameter in our study, and therefore we compare the \mhenv\ distribution of our progenitor models with that predicted for binary evolution using the Binary Population and Spectral Synthesis (\texttt{BPASS}) version 2.1 \citep{eldridge+17}.
At the same time, we included pre-SN single-star models calculated with a different code.
\citet{sukhbold+16} present single-star pre-SN models for a large \mzams\ range calculated with the \texttt{KEPLER} code \citep{weaver+78,woosley+02}, but here, the initial masses were limited to the 9$-$25~\ms\ range for consistency with the stellar models used in our work.
A similar range of values between our distribution and that from \citet{sukhbold+16} is found.
As before, the initial masses of primary stars (i.e. the initially more massive star of the system) of the \texttt{BPASS} models were restricted to the 9$-$25~\ms\ range. We also required that the star finish as a RSG.
Stars that lose mass through Roche-lobe overflow finish their evolution with smaller \mhenv\ than if they had evolved as single stars, producing a wider distribution that reach lower \mhenv\ values than our models.
If the effects of binary evolution are taken into account, significant diversity in hydrogen-rich envelope masses is found, from a minimal value of 0.1~\ms\ to a maximum value of 29~\ms\footnote{However, some of the stars in the \texttt{BPASS} distribution will not produce the \sneii\ studied here. Stars with low \mhenv\ might produce SN~IIb-like events. In addition, some stars can experience a merger and finish their evolution with \mhenv\,$\gtrsim$\,10~\ms, which might produce \sneii\ with \optd\ longer than $\sim$150~days \citep[called `long-SNe~IIP';][]{eldridge+18}.}.
Differences in the \mhenv\ distribution between single- and binary-star models do exist, with standard single-star evolution giving a narrower range of \mhenv\ that may affect our results.
However, the goal of this study is to test the predictions of standard single-star evolution against correlations between physical and observed properties of \sneii.
A study including the effects on binary evolution is warranted in the future.

%------------------------------------------------------
\subsection{\Ni\ mixing}
\label{sec:mixing}

Our analyses of the relative importance of each physical parameter for \stwo\ suggest that the explosion energy is the parameter that is driving  variation to the greatest degree, with \mix\ showing only a minor influence (see Table~\ref{table:kda}).
However, in Sect.~\ref{sec:corr_std} we find a strong anti-correlation between \mix\ and \stwo\ in the sense that fast-declining \sneii\ are consistent with a more concentrated \Ni.

The effect of \mix\ in shaping LCs of \sneii\ was previously studied in the literature. The degree of \mix\ within the envelope determines the time at which \Ni\ starts to affect the LC \citep[e.g.][]{bersten+11,kozyreva+19}.
While an extensively mixed \Ni\ influences the LC from the early plateau phase, a more concentrated \Ni\ distribution starts affecting the LC during the late recombination phase \citep[see][their Fig.~12]{bersten+11}.
Before the photosphere recedes to the shells containing \Ni, the behaviour of the LC is similar to that in the absence of \Ni, which naturally causes a decline during the plateau phase.
Therefore, \Ni\ mostly concentrated in the inner ejecta is necessary to produce rapid decline in the plateau phase (higher \stwo\ values).
In summary, we find that fast declining \sneii\ are consistent with higher energy explosions and small-scale \mix. We now analyse possible connections between both physical parameters.

The mixing of heavy elements through the ejecta is caused by Rayleigh$-$Taylor instabilities that appear once the shock wave passes through the composition interfaces between the carbon-oxygen core and the helium core, and this latter and the hydrogen-rich envelope.
Numerous numerical simulations have been carried out using blue supergiant progenitors to reproduce the large-scale mixing of \Ni\ presented in SN~1987A, but only a small number of explosion models with RSGs have been calculated.
Recently, \citet{wongwathanarat+15} computed 3D simulations and found that the degree of \mix\ depends on the progenitor structure (e.g. the size of the carbon-oxygen core, the density structure of the helium core, and the density gradient at the interface between the helium core and the hydrogen-rich envelope), the explosion energy, and the asymmetries created by the neutrino-driven mechanism.
More energetic explosions produce a faster propagation of \Ni\ within the ejecta, but also result in a faster shock wave \citep{wongwathanarat+15}. 
Therefore, in principle, the explosion energy is not expected to correlate with the mixing of \Ni, implying that the correlation found between \mix\ and the plateau declination rate is authentic.

The analysis of the effect of each physical parameter on the observed diversity of \sneii\ shows that the mixing of \Ni\ within the ejecta produces minor changes in \optd, although at the same time, a strong correlation is found between both parameters based on the results achieved with the CSP-I sample (see Sect.~\ref{sec:corr_std}).
Thus, this correlation may be a consequence of some additional correlations, such as: \sneii\ declining faster have shorter \pd\ \citep[e.g.][\citetalias{anderson+14_lc}; Fig.~\ref{fig:corr_matrix_obs_pars}]{pskovskii67}, and are consistent with more concentrated mixing of \Ni\ (Fig.~\ref{fig:corr_s2_vel_std}).

%------------------------------------------------------
\subsection{Observational biases}

The observed SN~II sample used in this work is of high quality in terms of LC cadence, wavelength coverage, and photometric accuracy. However, it still contains observational biases that may affect our results and conclusions. 
The CSP-I SN~II sample is magnitude-limited ---SNe were only chosen for followup if they were brighter than around 18th magnitude (at optical wavelengths) at maximum light. Thus, our sample is biased towards intrinsically brighter \sneii, meaning we lack lower luminosity events (as compared to their absolute rate of explosion). 
Figure~\ref{fig:corr_mbol_std} shows that low-luminosity events are almost exclusively constrained to arise from low-energy explosions, and therefore our sample is probably less populated with such explosions compared to the number that exist in nature. However, it is not clear that such a bias will affect our results on correlations between observed and physical parameters, or our conclusions on the dominant physical parameters driving SN~II diversity.

The magnitude-limited nature of our sample also means that we will be biased against heavily extinguished events. However, there is no evidence (to our knowledge) that the amount of host-galaxy extinction suffered by a SN~II correlates with any of its intrinsic physical properties, and therefore we do not believe that such a bias will affect our results. Finally, the CSP-I \snii\ sample was obtained at a time when most surveys discovering nearby \sneii\ (that would be candidates for followup by the CSP-I) were targeted in nature, with relatively small fields of view. Such surveys were biased against discovering \sneii\ in low-luminosity hosts. Galaxy luminosity correlates with galaxy metallicity, which probably means that our sample lacks SNe~II arising from low-metallicity progenitors. \citet{gutierrez+18} analysed SNe~II exploding in such dim galaxies, concluding that in general they had more slowly declining LCs, but otherwise had observational parameters similar to the rest of the SN~II population. As presented in Figure~\ref{fig:corr_s2_vel_std}, we find more slowly declining SNe~II (lower \stwo) to generally arise from lower energy explosions. Therefore, similarly to the above, the lack of SNe~II in low-luminosity hosts in our sample may have led to a lack of low-energy explosions. 
This may somewhat increase the uncertainties when making more general statements, but this should not significantly affect our results or conclusions for the sample as it is.

%------------------------------------------------------
\section{Conclusions}
\label{sec:conclusions}

In this study, we present an analysis of correlations between the observed and physical properties of a statistically significant sample of \sneii\ from the CSP-I in order to understand their diversity in terms of progenitor and explosion properties.
We used the physical parameters determined in \citetalias{martinez+21b} through hydrodynamical modelling of \snii\ bolometric LCs and expansion velocities, such as \mzams\ (which relates to \mej, \mhenv, and \ra), \e, \mni, and the distribution of \Ni\ within the ejecta, together with many photometric and spectroscopic parameters.

This study shows that the explosion energy is the physical parameter that correlates with the highest number of observed parameters, including \pd, \optd, \mbolend, \mboltail, \stwo, expansion velocities, the pEWs of some metal lines, and the flux ratio of the absorption to emission component of the \ha\ P-Cygni profile. In addition, we find that higher energy explosions manifest more \Ni\ in the ejecta.
Faster declining \sneii\ are consistent with higher energy explosions and more concentrated \Ni\ in the inner regions of the ejecta. In contrast to previous results, we find zero correlation between the declination rate during the plateau and the hydrogen-rich envelope mass. However, we note the caveat that only a narrow range of hydrogen-rich envelope masses was studied.
We measured parameters from our grid of bolometric LC and photospheric velocity models to determine the effect of each physical parameter on the observables through a statistical study. According to our findings, the explosion energy is the parameter causing the greatest impact on \snii\ diversity. 

New progenitor pre-SN structures evolved with enhanced mass loss were considered to model the observations of two short-plateau \sneii\ in the CSP-I sample: SN~2006Y and SN~2008bu. 
These models were also used to reproduce several other \sneii\ in the CSP-I sample successfully modelled in \citetalias{martinez+21b}, finding that nine additional objects are better reproduced with the non-standard models.
Furthermore, using these new results from non-standard stellar evolution, we find a significant increase in the strength of correlations between \mhenv\ and \pd, \mhenv\ and \optd, and \mej\ and \sthr. 
This implies that more mass loss is needed to reproduce short-plateau \sneii\  than is predicted by standard single-star
evolution\ (as noted by previous studies), but also that non-standard models are necessary for a complete understanding of \snii\ diversity. 
The differences in the strength of correlations when different pre-SN models are used clearly shows the impact of having different treatments of stellar evolution.

A significant number of \sneii\ decline faster during the radioactive tail phase (11 out of 16, see \citetalias{martinez+21a}) than the expected rate whilst assuming full trapping of gamma-ray photons. The moderate correlation found between \sthr\ and \mej\ ---when non-standard models are included--- implies that \sthr\ is directly affected by the SN ejecta mass in accordance with previous suggestions \citepalias{anderson+14_lc,gutierrez+17II}.

This paper concludes a three-part series using the CSP-I \snii\ sample to understand the SN~II phenomenon. Constraints are presented on progenitor and explosion properties through derivation of physical parameters from comparison between observed and modelled SNe. The current work further constrains the importance of different SN~II physical parameters in explaining SN~II diversity. However, these investigations also highlight limitations in our understanding of the lives and explosive deaths of massive stars, thus encouraging further modelling and additional high-quality observations.

%------------------------------------------------------
\begin{acknowledgements}
We thank the referee for the useful comments that improved the manuscript.
The work of the Carnegie Supernova Project was supported by the National Science Foundation under grants AST-0306969, AST-0607438, AST-1008343, AST-1613426, AST-1613472, and AST-1613455.
L.M. acknowledges support from a CONICET fellowship.
L.M. and M.O. acknowledge support from UNRN~PI2018~40B885 grant.
M.H. acknowledges support from the Hagler Institute of Advanced Study at Texas A\&M University.
S.G.G. acknowledges support by FCT under Project CRISP PTDC/FIS-AST-31546/2017 and~Project~No.~UIDB/00099/2020.
M.S. is supported by grants from the VILLUM FONDEN (grant number 28021) and the Independent Research Fund Denmark (IRFD; 8021-00170B).
F.F. acknowledges support from the National Agency for Research and Development (ANID) grants: BASAL Center of Mathematical Modelling AFB-170001, Ministry of Economy, Development, and Tourism’s Millennium Science Initiative through grant IC12009, awarded to the Millennium Institute of Astrophysics, and FONDECYT Regular \#1200710.
L.G. acknowledges financial support from the Spanish Ministry of Science, Innovation and Universities (MICIU) under the 2019 Ram\'on y Cajal program RYC2019-027683 and from the Spanish MICIU project PID2020-115253GA-I00.
P.H. acknowledges the support by National Science Foundation (NSF) grant AST-1715133.
This work made use of v2.2.1 of the Binary Population and Spectral Synthesis (BPASS) models as described in \citet{eldridge+17} and \citet{stanway+18}.
\\
\textit{Software:} \texttt{emcee} \citep{emcee}, \texttt{NumPy} \citep{numpyguide2006,numpy2011}, \texttt{matplotlib} \citep{matplotlib}, \texttt{MESA} \citep{paxton+11,paxton+13,paxton+15,paxton+18,paxton+19}, \texttt{SciPy} \citep{scipy2020}, \texttt{Kruskals}, \texttt{Pandas} \citep{pandas}, \texttt{ipython/jupyter} \citep{jupyter}.
\end{acknowledgements}

%------------------------------------------------------
\bibliographystyle{aa}
\bibliography{biblio}

%--------------------------------------------------
\onecolumn
\begin{appendix}
%--------------------------------------------------
\begin{multicols}{2}
\section{\snii\ diversity and its physical origin}
\label{appendix:plots}

\begin{figure*}[h]
\centering
\includegraphics[width=1.0\textwidth]{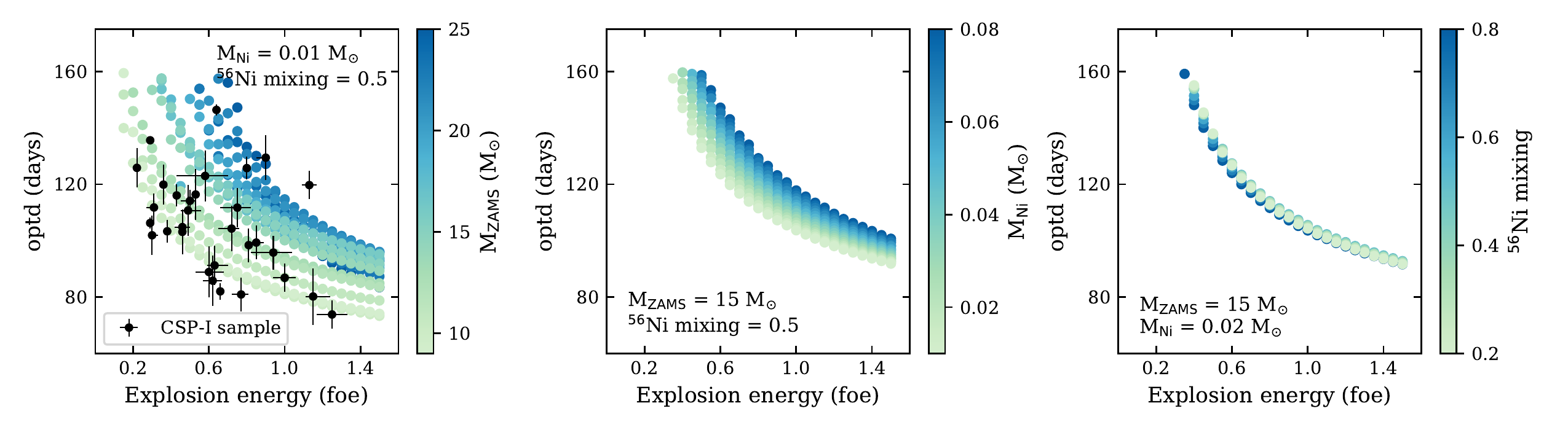}
\caption{Optically thick phase duration (\optd) measured from the synthetic bolometric LCs as a function of the explosion energy. The physical parameters not being varied are presented in each subplot together with their fixed values. Each subplot shows the influence of the other physical parameters: \mzams\ (left panel), \mni\ (middle panel), and \mix\ (right panel). Only models with \optd\ values smaller than 160~days are analysed. 
Black dots represent the observations from the CSP-I \snii\ sample. Some observations fall outside the range of the model parameters due to the fixed physical parameters. Changes in the fixed values represent different ranges of model parameters.}
\label{fig:optd}
\end{figure*}

\begin{figure*}
\centering
\includegraphics[width=1.0\textwidth]{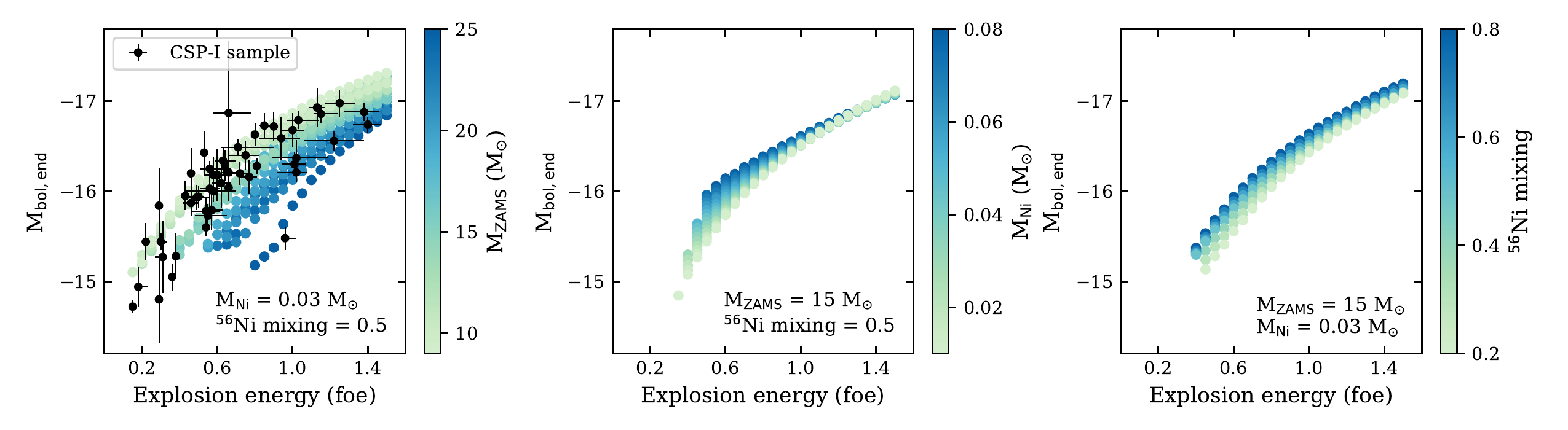}
\caption{\mbolend\ measured from the synthetic bolometric LCs as a function of the explosion energy. The physical parameters not being varied are presented in each subplot together with their fixed values. Each subplot shows the influence of the other physical parameters: \mzams\ (left panel), \mni\ (middle panel), and \mix\ (right panel). Only models with \optd\ values smaller than 160~days are analysed. 
Black dots represent the observations from the CSP-I \snii\ sample. Some observations fall outside the range of the model parameters due to the fixed physical parameters. Changes in the fixed values represent different ranges of model parameters.}
\label{fig:mbolend}
\end{figure*}

\begin{figure*}
\centering
\includegraphics[width=1.0\textwidth]{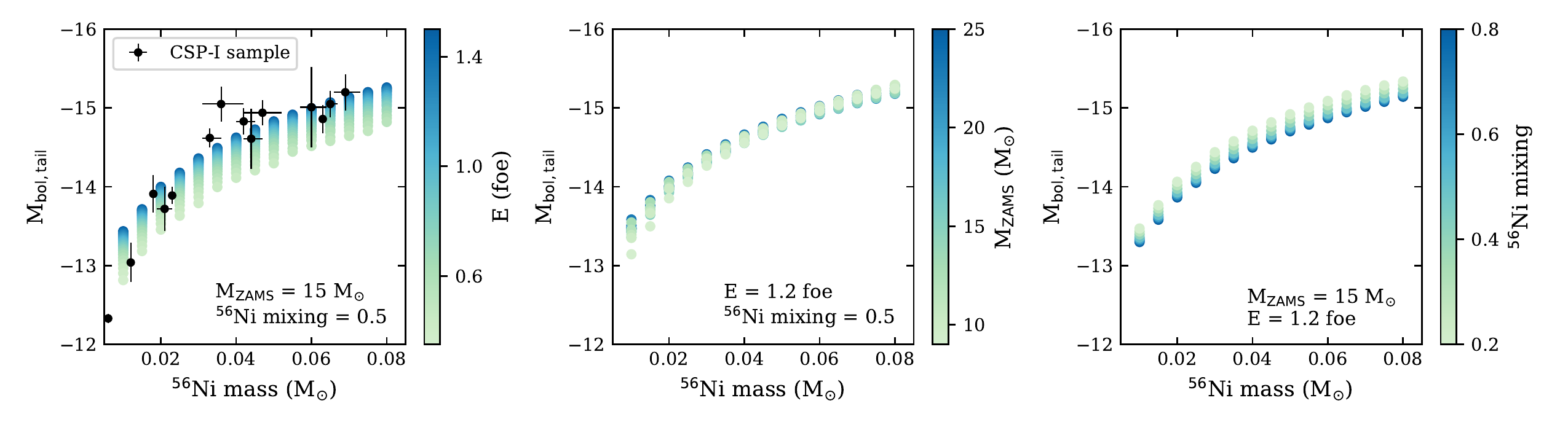}
\caption{\mboltail\ measured from the synthetic bolometric LCs as a function of \mni. The physical parameters not being varied are presented in each subplot together with their fixed values. Each subplot shows the influence of the other physical parameters: \e\ (left panel), \mzams\ (middle panel), and \mix\ (right panel). Only models with \optd\ values smaller than 160~days are analysed.
Black dots represent the observations from the CSP-I \snii\ sample. Some observations fall outside the range of the model parameters due to the fixed physical parameters. Changes in the fixed values represent different ranges of model parameters.}
\label{fig:mboltail}
\end{figure*}

\begin{figure*}
\centering
\includegraphics[width=1.0\textwidth]{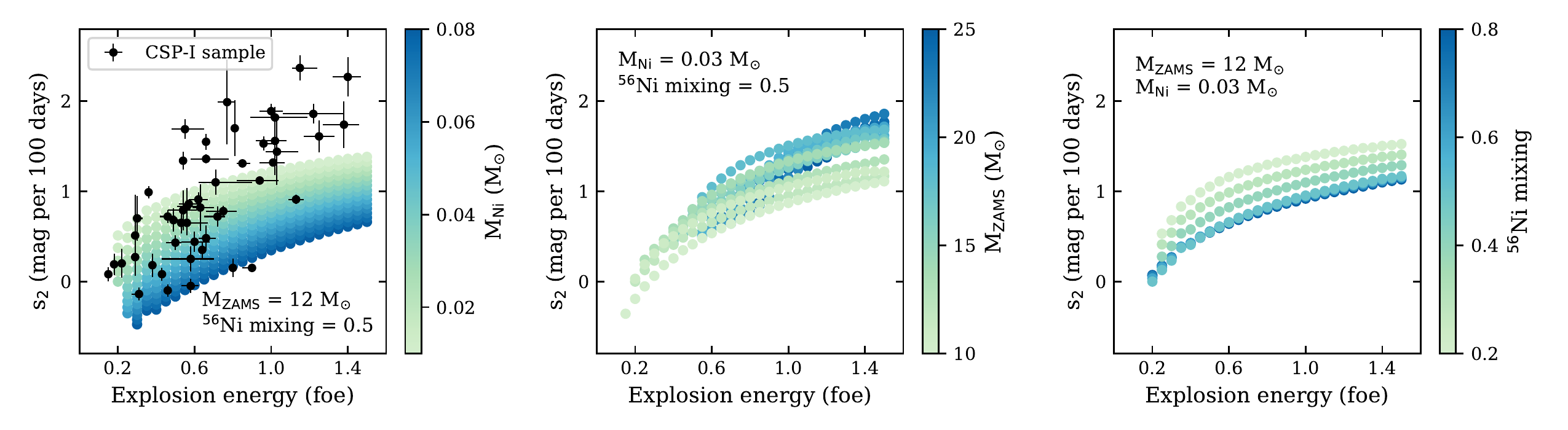}
\caption{Declination rate during the plateau (\stwo) measured from the synthetic bolometric LCs as a function of explosion energy. The physical parameters not being varied are presented in each subplot together with their fixed values. Each subplot shows the influence of the other physical parameters: \mni\ (left panel), \mzams\ (middle panel), and \mix\ (right panel). Only models with \optd\ values smaller than 160~days are analysed. 
Black dots represent the observations from the CSP-I \snii\ sample. Some observations fall outside the range of the model parameters due to the fixed physical parameters. Changes in the fixed values represent different ranges of model parameters.}
\label{fig:s2}
\end{figure*}

\begin{figure*}
\centering
\includegraphics[width=1.0\textwidth]{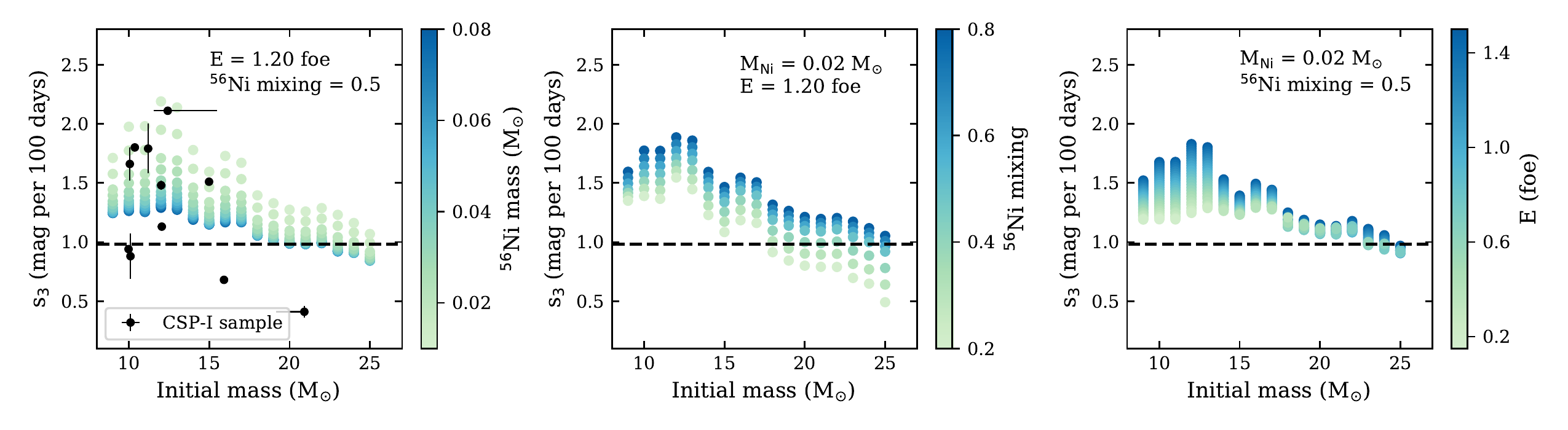}
\caption{Declination rate of the radioactive tail (\sthr) measured from the synthetic bolometric LCs as a function of \mzams. The physical parameters not being varied are presented in each subplot together with their fixed values. Each subplot shows the influence of the other physical parameters: \mni\ (left panel), \mix\ (middle panel), and \e\ (right panel). The horizontal dashed lines indicate the expected declination rate for full trapping of emission from $^{56}$Co decay. Only models with \optd\ values smaller than 160~days are analysed.
Black dots represent the observations from the CSP-I \snii\ sample. Some observations fall outside the range of the model parameters due to the fixed physical parameters. Changes in the fixed values represent different ranges of model parameters.}
\label{fig:s3}
\end{figure*}

\begin{figure*}
\centering
\includegraphics[width=1.0\textwidth]{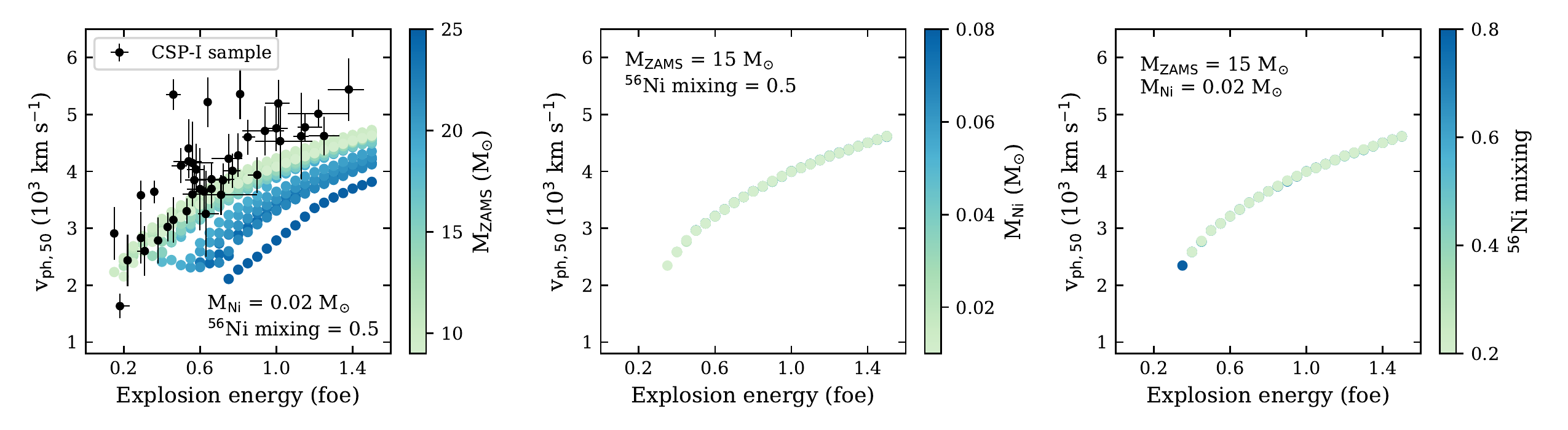}
\caption{Photospheric velocity at 50~days measured from the models (\vphfi) as a function of explosion energy. The physical parameters not being varied are presented in each subplot together with their fixed values. Each subplot shows the influence of the other physical parameters: \mzams\ (left panel), \mni\ (middle panel), and \mix\ (right panel). Only models with \optd\ values smaller than 160~days are analysed. Black dots represent the observations from the CSP-I \snii\ sample. Some observations fall outside the range of the models (see text).}
\label{fig:v50}
\end{figure*}

In Sect.~\ref{sec:corr_std} we present an analysis of correlation between physical and observed parameters for the CSP-I \snii\ sample.
In addition, with the aim of determining the effect of physical parameters on \snii\ diversity, we measured some parameters directly from our grid of bolometric LC and photospheric velocity models (Sect.~\ref{sec:models}).
Here, in Figs.~\ref{fig:optd} to \ref{fig:v50}, we show observables measured from our grid of explosion models against the physical parameter yielding the highest relative importance, following the results from Table~\ref{table:kda}. Each panel presents the effect of a distinct physical parameter denoted by different colours given in the colour bar. The physical parameters not being varied are presented in each panel together with their fixed values.

Figure~\ref{fig:optd} shows that the large range of \optd\ values seen in the previous sections are mainly produced by \mzams\ and \e, in line with previous theoretical studies. 
The left panel clearly shows the great dependency of these two physical parameters on the \optd, while middle and right panels show the effect of \mni\ and its mixing in the \e$-$\optd\ relation, respectively.
As expected, higher \mni\ extends the optically thick phase, although \Ni\ distribution in the ejecta does not seem to produce significant changes. 

The brightness at the end of the plateau (\mbolend) is highly affected by the explosion energy, with higher energy explosions delivering more luminous \sneii\ (Fig.~\ref{fig:mbolend}). 
\mzams\ is the physical parameter that produces the largest deviation in the \e$-$\mbolend\ relation (Fig.~\ref{fig:mbolend}, left panel), while \mni\ and its degree of mixing within the ejecta alter this relation on smaller scales.
However, \mni\ has its major effect on \mbolend\ in the low-\e\ regime where \sneii\ are fainter and the effects of
\columnbreak
$^{56}$Co decay are more significant during the late photospheric phase, causing higher luminosities at late-plateau phases (Fig.~\ref{fig:mbolend}, middle panel).
Figure~\ref{fig:mboltail} indicates that \mboltail\ is mostly affected by the amount of \Ni\ in the ejecta. The explosion energy and the mixing of \Ni\ do not significantly contribute to \mboltail.

The explosion energy produces the largest effect on the \stwo\ declination rate, followed by \mni, at least with the models used in this series of papers (Fig.~\ref{fig:s2}).
\mzams\ also shows variations on \stwo, but with a lower effect than the explosion energy.
However, this analysis only includes standard pre-SN models, where none of the progenitors were evolved with significant mass loss.
The relation between \e\ and \stwo\ has an additional contribution to the dispersion produced by the mixing of \Ni, with faster declining \sneii\ consistent with low degrees of \mix.

The declination rate of the radioactive tail phase (\sthr) takes the theoretical value of 0.98~mag per 100 days assuming full trapping of the gamma-ray emission from $^{56}$Co decay (dashed lines in Fig.~\ref{fig:s3}).
However, a considerable number of \sneii\ are found with higher \sthr\ \citepalias{anderson+14_lc,gutierrez+17II,martinez+21a}.
Here, we find that \mzams\ is the main driver affecting the \sthr\ declination rate (Table~\ref{table:kda}, Fig.~\ref{fig:s3}). At least within the range of initial masses of our standard pre-SN models, high-\mzams\ stars imply higher \mej. Therefore, the high \sthr\ values are consistent with lower ejecta masses.
Low ejecta masses are less efficient in thermalising the gamma-ray emission from $^{56}$Co decay, leading to a more rapid decline in the luminosity during the radioactive tail phase.
The degree of \Ni\ mixing within the ejecta also contributes to the differences found in \sthr. This is seen in the middle panel of Fig.~\ref{fig:s3} where more extended \mix\ produces a more rapid declination of \sneii\ during the radioactive tail.
Figure~\ref{fig:v50} shows that \e\ and \mzams\ are the only two physical parameters that produce different photospheric velocities at 50~days from explosion. 
\end{multicols}

\begin{figure}
\includegraphics[width=0.38\textwidth]{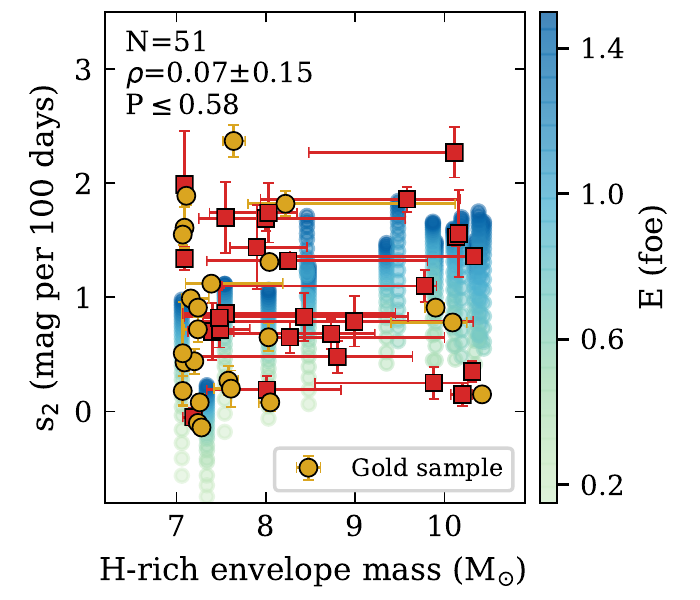}
\caption{Same as in Fig.~\ref{fig:corr_s2_vel_std} but for \mhenv\ versus \stwo.}
\label{fig:corr_hmass_s2_std}
\end{figure}

\FloatBarrier

%------------------------------------------------------
\begin{multicols}{2}
\section{Photometric and spectroscopic correlations for the CSP-I \snii\ sample}
\label{sec:corr_obs_pars}

In the main text of this paper, we present our analysis of correlations between physical and observed parameters, without discussing the relations between observed \snii\ parameters.
For completeness, in this Appendix we present correlations between bolometric LC parameters, colours at different epochs, and expansion velocities and pEWs measured at 50 days post-explosion.
Similar analyses have been carried out in the literature \citepalias{anderson+14_lc,gutierrez+17II,dejaeger+18}.
The only difference is that in those works the authors used $V$-band LC properties in their analyses, while here we used the bolometric LC parameters measured in \citetalias{martinez+21a}. 

\begin{figure*}[hb]
\centering
\includegraphics[width=0.87\textwidth]{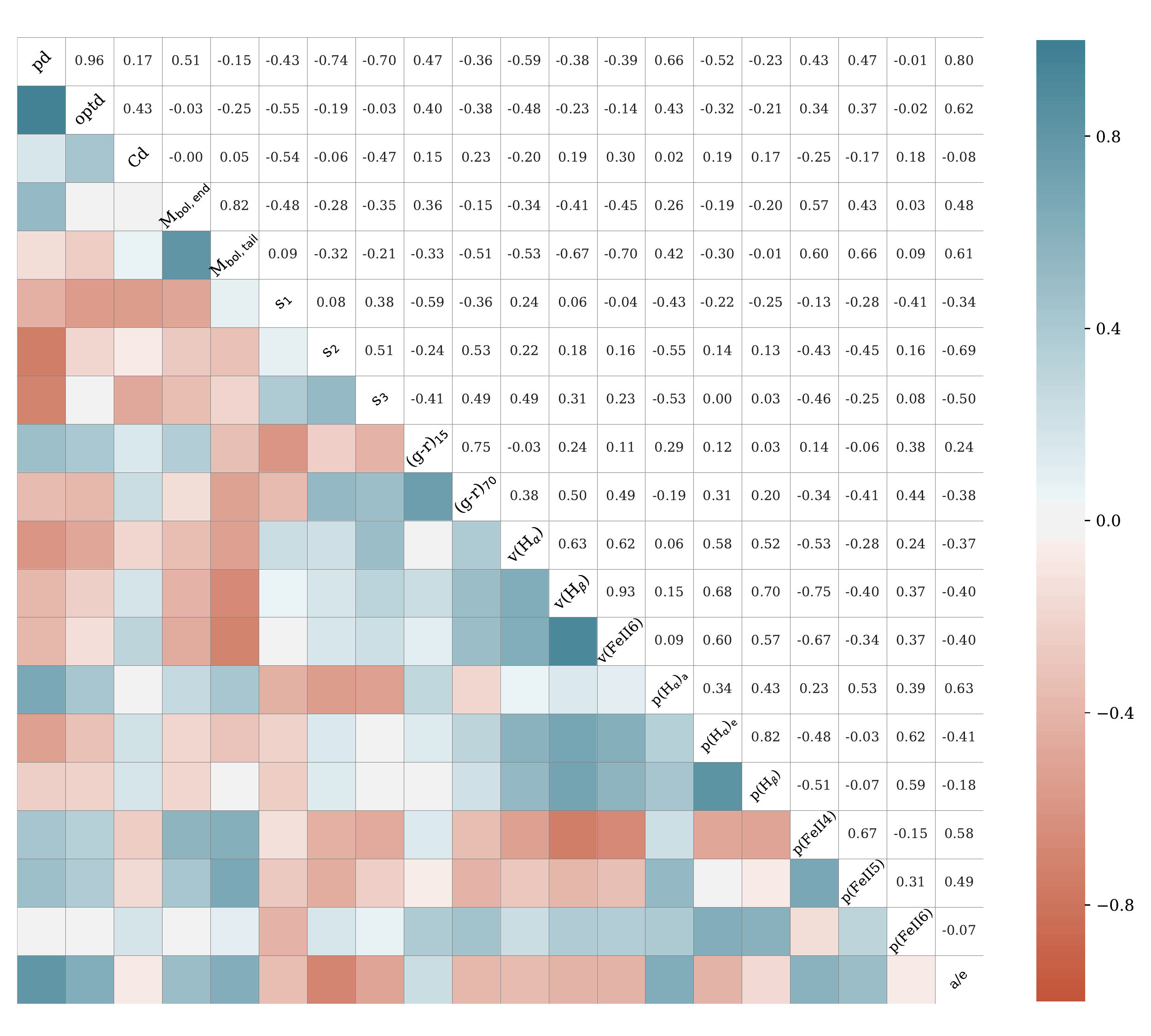}
\caption{Correlation matrix of the bolometric LC, spectral, and colour parameters used in the current study. The Pearson correlation coefficients are presented in the upper triangle, while in the lower triangle the correlation coefficients are colour-coded. The diagonal middle line shows the name of the parameters used: \pd, \optd, \cd, \mbolend, \mboltail, \sone, \stwo, \sthr, (g$-$r) at 15 and 70 days after explosion, line velocity of \ha, \hb, \ion{Fe}{ii} $\lambda$5169, pEW(\ha) of absorption component, pEW(\ha) of emission component, pEW(\hb), pEW(\ion{Fe}{ii} $\lambda$4924), pEW(\ion{Fe}{ii} $\lambda$5018), pEW(\ion{Fe}{ii} $\lambda$5169), and \aeha.}
\label{fig:corr_matrix_obs_pars}
\end{figure*}

Figure~\ref{fig:corr_matrix_obs_pars} shows the correlation matrix of the parameters.
Most of the trends found in previous studies are recovered with similar or higher degree of correlation. Similar to \citetalias{gutierrez+17II}, we find that \sneii\ with 
\columnbreak
shorter \pd\ are brighter, have faster declining LCs at the three measured epochs, lower pEW(\ha) of the absorption component and \aeha, and higher expansion velocities.
In addition, similar to \citetalias{dejaeger+18}, we find that fast-declining \sneii\ are bluer at early epochs but redder at later epochs.
Given that similar results are obtained, we do not go into the details of the correlations and their possible explanations. The reader is referred to \citetalias{anderson+14_lc}, \citetalias{gutierrez+17II}, and \citetalias{dejaeger+18} for further details.

While we recover the trend between \pd\ and \sthr\ (\p\,=\,$-$0.70\,$\pm$\,0.25, $N$\,=\,6), we find zero correlation between \optd\ and \sthr.
The lack of correlation between \sthr\ and \optd\ may be possible because, by definition, \optd\ includes the plateau phase, mostly related to the hydrogen-rich envelope mass, and other phases that are powered by different mechanisms (the release of shock deposited energy, ejecta-CSM interaction) that are related to different progenitor properties. However, \citetalias{gutierrez+17II} found a moderate correlation between \optd\ and \sthr.
\end{multicols}

\FloatBarrier

%--------------------------------------------------
\begin{multicols}{2}
\section{Physical parameters of \sneii\ using non-standard pre-SN models}
\label{app:non_std}

\begin{figure*}[hb]
\centering
\includegraphics[width=1.0\textwidth]{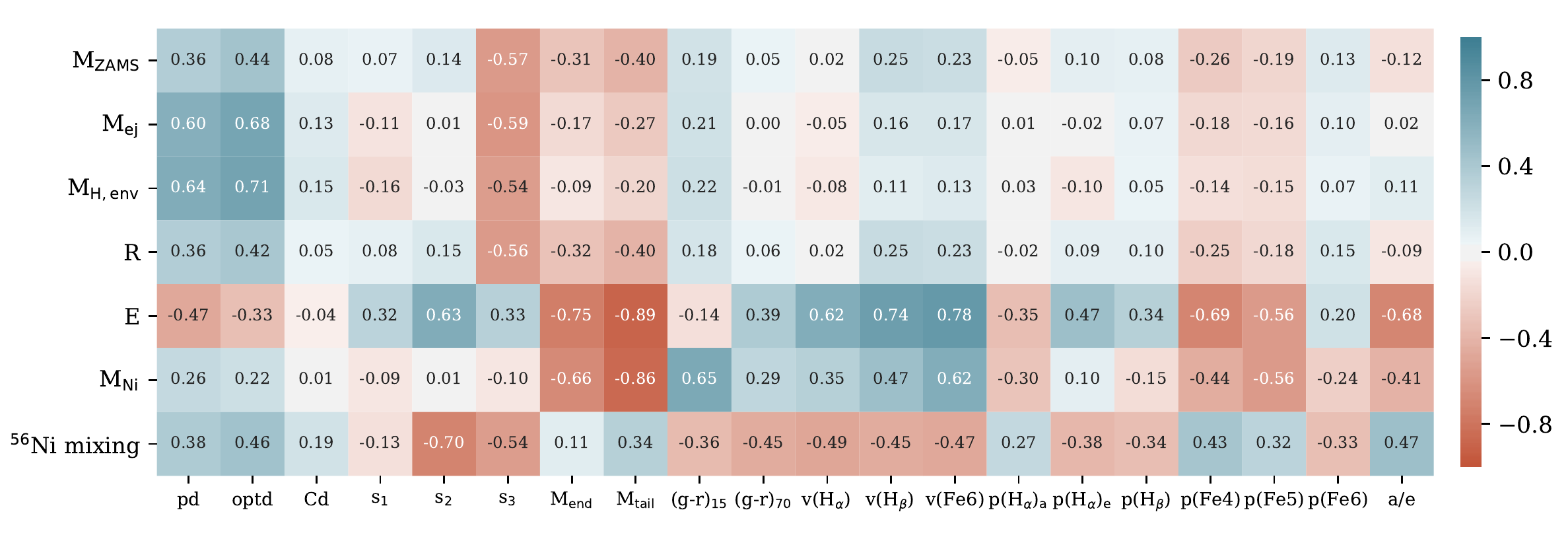}
\caption{Correlation matrix of the observed \snii\ parameters against the physical parameters using non-standard pre-SN models. For each pair, the Pearson correlation coefficient is given and colour-coded. The observed parameters shown are: \pd, \optd, \cd, \sone, \stwo, \sthr, \mbolend, \mboltail, \gr$_{15}$, \gr$_{70}$, velocity of \ha, \hb, and \ion{Fe}{ii} $\lambda$5169, pEW(\ha) of absorption component, pEW(\ha) of emission component, pEW(\hb), pEW(\ion{Fe}{ii} $\lambda$4924), pEW(\ion{Fe}{ii} $\lambda$5018), pEW(\ion{Fe}{ii} $\lambda$5169), and \aeha.}
\label{fig:corr_matrix_physical_observed_nonstd}
\end{figure*}

\begin{figure*}
\centering
\includegraphics[width=0.49\textwidth]{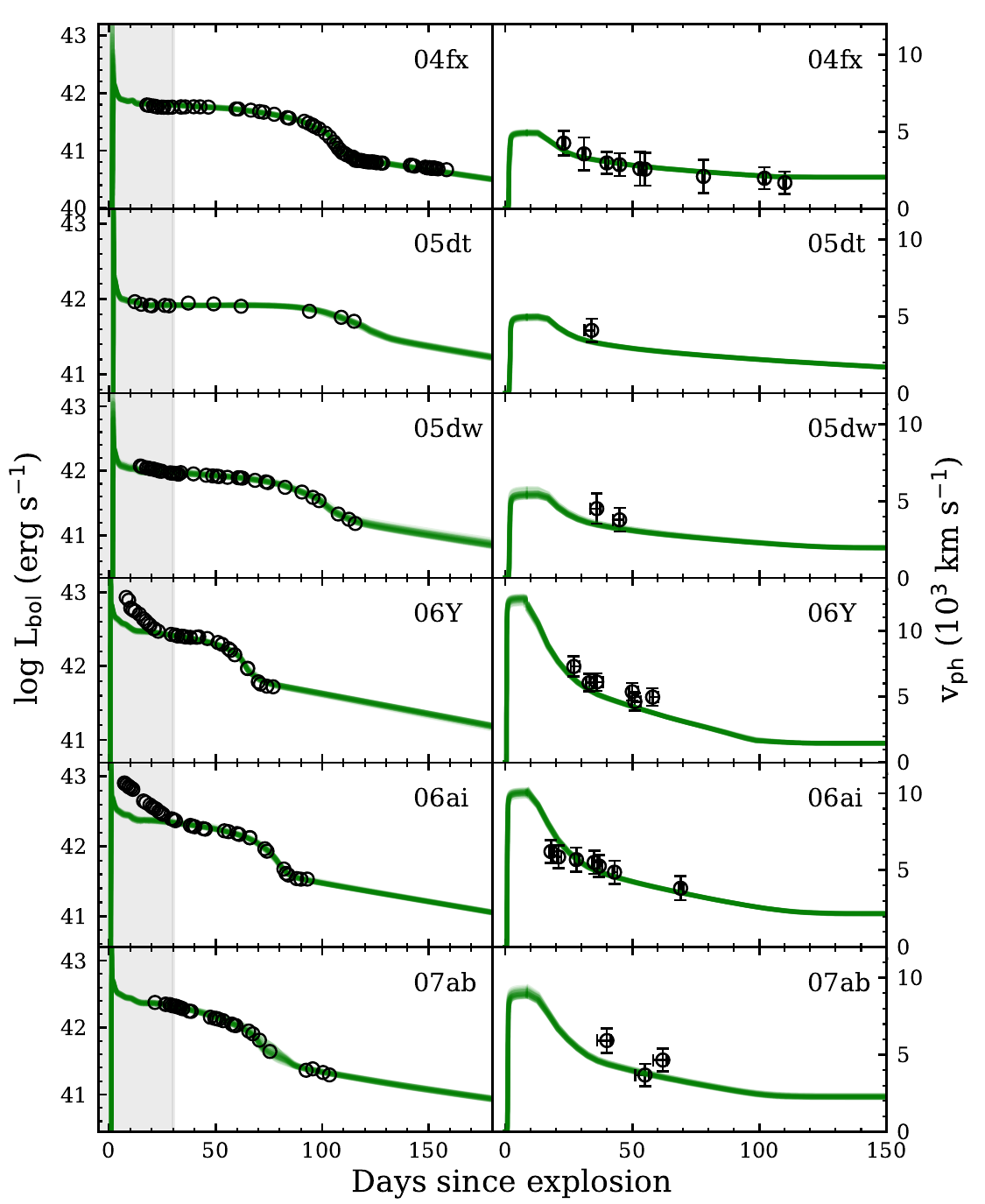} 
\includegraphics[width=0.49\textwidth]{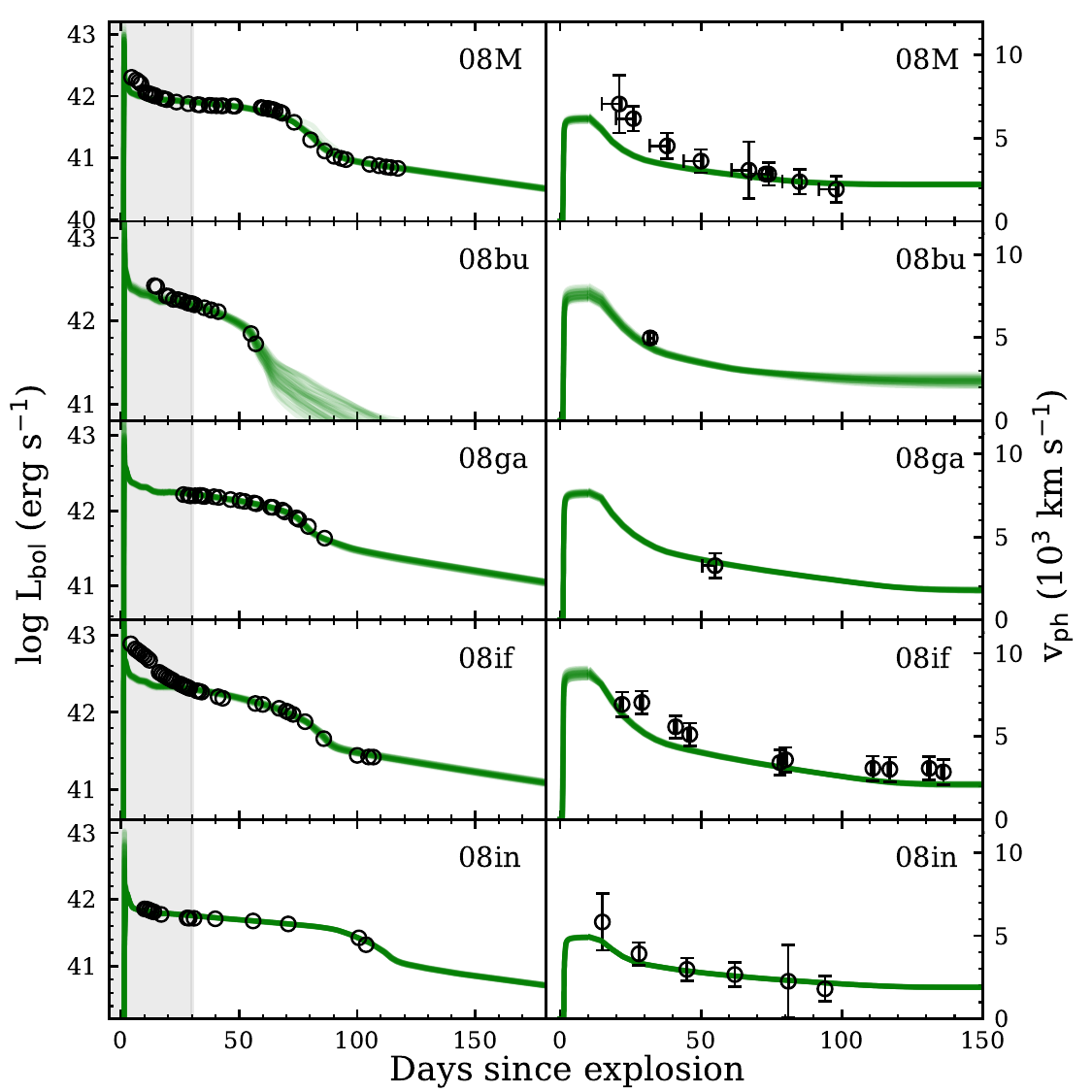}
\caption{Comparison between models and observations. The open circles show the observed bolometric LCs and \ion{Fe}{ii} velocities. Solid lines represent 30 models randomly chosen from the probability distribution. We used the models with incremented mass loss. The panels present SNe in order of their discovery dates. Grey shaded regions show the early data we removed from the fitting. The errors in the observed bolometric LCs are not plotted for better visualisation.}
\label{fig:fits}
\end{figure*}

Table~\ref{table:results_nonstd} presents the physical parameters derived from the hydrodynamical modelling of bolometric LCs and expansion velocities for the 11 \sneii\ modelled with non-standard pre-SN models. This table also includes values for \mej, \mhenv, \mh, and \ra. These quantities were not derived from the modelling, but were interpolated from the \mzams\ and $\eta$ values determined from the fitting.
The correlation matrix between observed and physical parameters using non-standard pre-SN models is presented in Fig.~\ref{fig:corr_matrix_physical_observed_nonstd}.
Figure~\ref{fig:fits} compares observations with models drawn from the posterior distribution of the parameters for the 11~\sneii\ modelled with non-standard pre-SN models.
\end{multicols}

\input{results_nonstd.tab}

\end{appendix}
\end{document}

%% file: physical_pars.tab
\begin{table}
\caption{Physical parameters for the grid of hydrodynamical models.}
\label{tab:models}
\centering
\resizebox{0.5\textwidth}{!}{
\begin{tabular}{cccc}
\hline\hline\noalign{\smallskip}
Parameter & Range & Increment & Extra values \\
\hline\noalign{\smallskip}
\mzams\ & 9$-$25~\ms\ & 1~\ms\ & --- \\
\e\ & 0.1$-$1.5~foe\tablefootmark{a},\tablefootmark{b} & 0.1~foe & --- \\
\mni\ & 0.01$-$0.08~\ms\ & 0.01~\ms\ & 10$^{-4}$, 5$\times$10$^{-3}$~\ms\ \\
\mix\ & 0.2$-$0.8\tablefootmark{c} & 0.3 & --- \\
\hline
\end{tabular}}
\tablefoot{
\tablefoottext{a}{1~foe~$\equiv$~10$^{51}$~erg.}
\tablefoottext{b}{Models for the largest masses and lowest energies were not computed due to numerical difficulties \citepalias[see][for details]{martinez+20}.}
\tablefoottext{c}{Given in fraction of the pre-SN mass.}
}
\end{table}

%% file: kda.tab
\begin{table}
    \caption{Relative importance of the physical parameters for each observable measured in our LC and photospheric velocity models.}
    \label{table:kda}
    \centering          
    \begin{tabular}{lcccc}
    \hline\hline\noalign{\smallskip}
     & $M_{\rm ZAMS}$ & \emph{E} & $M_{\rm Ni}$ & $^{56}$Ni mixing \\
    \hline\noalign{\smallskip}
    $optd$ & 0.28 & 0.66 & 0.06 & 0.00 \\
    $M_{\rm bol,end}$ & 0.22 & 0.69 & 0.03 & 0.06 \\
    $M_{\rm bol,tail}$ & 0.01 & 0.22 & 0.74 & 0.03 \\
    $s_{\rm 2}$ & 0.21 & 0.50 & 0.15 & 0.14 \\
    $s_{\rm 3}$ & 0.58 & 0.03 & 0.20 & 0.19 \\
    \it{v}$_{\rm ph,50}$ & 0.33 & 0.67 & 0.00 & 0.00 \\
    \hline
    \end{tabular}
\end{table}

%% file: presn_models.tab
\begin{table}
\centering
\caption{Progenitor properties for the pre-SN models evolved with enhanced mass loss and the standard pre-SN models.}
\label{table:presn_models}
\medskip
\begin{tabular}{lcccccc}
    \hline\hline\noalign{\smallskip}
    \mzams\ & $\eta$ & $M_{\rm presn}$ & \ra\ & \mhenv\ & \mh\ & $M_{\rm He}$ \\
    ($M_{\odot}$) & & ($M_{\odot}$) & ($R_{\odot}$) & ($M_{\odot}$) & ($M_{\odot}$) & ($M_{\odot}$) \\
    \hline\noalign{\smallskip}
    10 & 1.0 & 9.53 & 462 & 7.06 & 4.79 & 2.47 \\
    10 & 3.0 & 8.56 & 476 & 6.22 & 4.21 & 2.34 \\
    10 & 6.0 & 7.32 & 472 & 5.08 & 3.40 & 2.24 \\
    10 & 9.0 & 6.28 & 455 & 4.28 & 2.74 & 2.00 \\
    12 & 1.0 & 11.08 & 594 & 8.03 & 5.42 & 3.05 \\
    12 & 3.0 & 9.24 & 627 & 6.21 & 4.16 & 3.03 \\
    12 & 5.0 & 7.78 & 618 & 4.86 & 3.24 & 2.92 \\
    12 & 7.0 & 6.38 & 635 & 3.53 & 2.33 & 2.85 \\
    14 & 1.0 & 13.19 & 742 & 9.35 & 6.12 & 3.84 \\
    14 & 2.5 & 10.01 & 817 & 6.15 & 4.12 & 3.86 \\
    14 & 3.5 & 8.49 & 841 & 4.69 & 3.11 & 3.80 \\
    14 & 4.0 & 7.74 & 851 & 3.96 & 2.62 & 3.78 \\
    \hline
\end{tabular}
\tablefoot{The table includes the initial mass (\mzams), the wind scaling factor ($\eta$), the final mass at core collapse ($M_{\rm presn}$), the progenitor radius (\ra), the hydrogen-rich envelope mass (\mhenv), the total mass of hydrogen (\mh), and the helium-core mass ($M_{\rm He}$).}
\end{table}

%% file: corr_comparison.tab
\begin{table}
\centering
\caption{Comparison of the strength of correlations ($\rho$) with the largest differences between standard (std) and non-standard (non-std) results.}
\label{table:corr_comparison}
\medskip
\resizebox{0.49\textwidth}{!}{
\begin{tabular}{lccc}
\hline\hline\noalign{\smallskip}
Correlation & $N$ (std) & $\rho$ (std) & $P$ (std) \\
            & $N$ (non-std) & $\rho$ (non-std) & $P$ (non-std) \\
\hline\noalign{\smallskip}
$M_{\rm H,env}-pd$ & 20 & 0.43~$\pm$~0.15 & $\leq$0.23 \\
                           & 22 & 0.64~$\pm$~0.10 & $\leq$9.2\,$\times$\,10$^{-3}$ \\
\hline\noalign{\smallskip}
$M_{\rm H,env}-optd$ & 31 & 0.60~$\pm$~0.11 & $\leq$5\,$\times$\,10$^{-3}$ \\
                           & 33 & 0.71~$\pm$~0.09 & $\leq$1.1\,$\times$\,10$^{-4}$ \\
\hline\noalign{\smallskip}
$M_{\rm ej}-s_{3}$ & 11 & $-$0.47~$\pm$~0.32 & $\leq$0.66 \\
                   & 12 & $-$0.59~$\pm$~0.24 & $\leq$0.26 \\
\hline\noalign{\smallskip}
$M_{\rm Ni}-pd$ & 6 & 0.70~$\pm$~0.36 & $\leq$0.50 \\
                & 7 & 0.26~$\pm$~0.48 & $\leq$0.63 \\
\hline\noalign{\smallskip}
$M_{\rm Ni}-optd$ & 15 & 0.49~$\pm$~0.21 & $\leq$0.31 \\
                  & 16 & 0.22~$\pm$~0.31 & $\leq$0.74 \\
\hline
\end{tabular}}
\tablefoot{Columns: (1) correlation; (2) number of SNe~II within that correlation; (3) correlation coefficient; (4) upper limit of the significance of correlations.}
\end{table}

%% file: results_nonstd.tab
\begin{table*}[hb]
\centering
\caption{SN~II physical parameters using non-standard pre-SN models.}
\label{table:results_nonstd}
\medskip
\resizebox{1.0\textwidth}{!}{
\begin{tabular}{lccccccccccc}
\hline\hline\noalign{\smallskip}
SN & $t_{\rm exp}$ & scale & $M_{\rm ZAMS}$ & $\eta$ & $M_{\rm ej}$ & $M_{\rm H,env}$ & $M_{\rm H}$ & \emph{R} & $\emph{E}$ & $M_{\rm Ni}$ & $^{56}\rm Ni$ mixing \\
 & (MJD) & & ($M_{\odot}$) & & ($M_{\odot}$) & ($M_{\odot}$) & ($M_{\odot}$) & ($R_{\odot}$) & (foe) & ($M_{\odot}$) & \\
\hline\noalign{\smallskip}
2006Y & 53767.0$^{+2.3}_{-1.2}$ & 1.26$^{+0.01}_{-0.01}$ & 10.28$^{+0.72}_{-0.20}$ & 8.42$^{+0.25}_{-0.81}$ & 5.49$^{+0.27}_{-0.62}$ & 4.44$^{+0.21}_{-0.92}$ & 2.87$^{+0.17}_{-0.54}$ & 458$^{+178}_{-1}$ & 1.18$^{+0.05}_{-0.35}$ & 0.075$^{+0.003}_{-0.005}$ & 0.51$^{+0.10}_{-0.04}$ \\
\noalign{\smallskip}
2008bu & 54561.1$^{+1.2}_{-0.9}$ & 1.03$^{+0.17}_{-0.08}$ & 11.52$^{+0.61}_{-0.40}$ & 6.56$^{+0.68}_{-0.63}$ & 5.20$^{+0.42}_{-0.33}$ & 3.83$^{+0.40}_{-0.31}$ & 2.54$^{+0.27}_{-0.21}$ & 632$^{+4}_{-5}$ & 0.52$^{+0.10}_{-0.05}$ & --- & 0.32$^{+0.19}_{-0.09}$ \\
\noalign{\smallskip}
\hline
\noalign{\smallskip}
2004fx & 53302.6$^{+0.6}_{-0.6}$ & 1.17$^{+0.06}_{-0.06}$ & 10.35$^{+0.41}_{-0.25}$ & 3.15$^{+0.17}_{-0.11}$ & 7.48$^{+0.04}_{-0.07}$ & 6.15$^{+0.04}_{-0.06}$ & 4.16$^{+0.04}_{-0.06}$ & 477$^{+1}_{-1}$ & 0.30$^{+0.01}_{-0.01}$ & 0.011$^{+0.001}_{-0.001}$ & 0.78$^{+0.02}_{-0.04}$ \\
\noalign{\smallskip}
2005dt & 53613.1$^{+1.2}_{-2.9}$ & 0.95$^{+0.04}_{-0.02}$ & 12.25$^{+0.53}_{-0.70}$ & 3.33$^{+0.29}_{-0.25}$ & 7.47$^{+0.15}_{-0.20}$ & 5.96$^{+0.18}_{-0.18}$ & 3.99$^{+0.13}_{-0.13}$ & 626$^{+1}_{-31}$ & 0.33$^{+0.02}_{-0.02}$ & --- & 0.70$^{+0.04}_{-0.04}$ \\
\noalign{\smallskip}
2005dw & 53610.5$^{+1.5}_{-2.6}$ & 0.95$^{+0.07}_{-0.05}$ & 11.94$^{+0.76}_{-0.93}$ & 3.31$^{+0.14}_{-0.12}$ & 7.49$^{+0.08}_{-0.10}$ & 5.99$^{+0.15}_{-0.09}$ & 4.01$^{+0.12}_{-0.07}$ & 622$^{+4}_{-65}$ & 0.38$^{+0.05}_{-0.02}$ & --- & 0.70$^{+0.07}_{-0.09}$ \\
\noalign{\smallskip}
2006ai & 53778.9$^{+1.1}_{-0.7}$ & 0.95$^{+0.12}_{-0.04}$ & 10.60$^{+0.68}_{-0.42}$ & 3.45$^{+0.24}_{-0.22}$ & 7.36$^{+0.09}_{-0.04}$ & 6.04$^{+0.08}_{-0.17}$ & 4.08$^{+0.06}_{-0.14}$ & 476$^{+109}_{-1}$ & 1.04$^{+0.07}_{-0.03}$ & 0.047$^{+0.007}_{-0.003}$ & 0.55$^{+0.13}_{-0.07}$ \\
\noalign{\smallskip}
2007ab & 54131.9$^{+1.5}_{-3.8}$ & 0.97$^{+0.06}_{-0.04}$ & 11.82$^{+0.79}_{-0.80}$ & 4.31$^{+0.63}_{-0.78}$ & 6.82$^{+0.60}_{-0.47}$ & 5.37$^{+0.62}_{-0.43}$ & 3.59$^{+0.44}_{-0.30}$ & 617$^{+2}_{-52}$ & 0.84$^{+0.16}_{-0.10}$ & 0.031$^{+0.002}_{-0.001}$ & 0.31$^{+0.16}_{-0.07}$ \\
\noalign{\smallskip}
2008M & 54479.7$^{+0.8}_{-6.2}$ & 0.84$^{+0.05}_{-0.06}$ & 10.31$^{+0.35}_{-0.22}$ & 8.10$^{+0.47}_{-2.15}$ & 5.59$^{+0.76}_{-0.16}$ & 4.52$^{+0.59}_{-0.12}$ & 2.94$^{+0.48}_{-0.10}$ & 460$^{+12}_{-3}$ & 0.34$^{+0.03}_{-0.02}$ & 0.013$^{+0.001}_{-0.001}$ & 0.78$^{+0.02}_{-0.04}$ \\
\noalign{\smallskip}
2008ga & 54717.9$^{+0.4}_{-4.6}$ & 1.25$^{+0.03}_{-0.09}$ & 11.82$^{+0.67}_{-0.90}$ & 4.79$^{+0.16}_{-0.87}$ & 6.47$^{+0.69}_{-0.12}$ & 5.05$^{+0.81}_{-0.11}$ & 3.36$^{+0.59}_{-0.08}$ & 619$^{+1}_{-143}$ & 0.58$^{+0.11}_{-0.02}$ & --- & 0.69$^{+0.05}_{-0.10}$ \\
\noalign{\smallskip}
2008if & 54812.2$^{+0.4}_{-0.7}$ & 0.97$^{+0.07}_{-0.06}$ & 11.51$^{+0.99}_{-0.63}$ & 3.20$^{+0.21}_{-0.14}$ & 7.57$^{+0.01}_{-0.15}$ & 6.09$^{+0.09}_{-0.17}$ & 4.09$^{+0.10}_{-0.13}$ & 593$^{+33}_{-117}$ & 0.87$^{+0.04}_{-0.04}$ & 0.041$^{+0.004}_{-0.003}$ & 0.35$^{+0.07}_{-0.07}$ \\
\noalign{\smallskip}
2008in & 54827.4$^{+0.0}_{-0.0}$ & 1.63$^{+0.01}_{-0.01}$ & 10.05$^{+0.08}_{-0.03}$ & 3.64$^{+0.11}_{-0.08}$ & 7.28$^{+0.03}_{-0.05}$ & 5.97$^{+0.03}_{-0.04}$ & 4.03$^{+0.02}_{-0.03}$ & 476$^{+1}_{-1}$ & 0.29$^{+0.01}_{-0.01}$ & --- & 0.52$^{+0.03}_{-0.03}$ \\
\noalign{\smallskip}
\hline
\end{tabular}}
\tablefoot{Results are characterised by the median of the marginal distributions, adopting the 16th and 84th percentiles as the lower and upper uncertainties. Columns: (1) SN name; (2) explosion epoch; (3) scale factor; (4) progenitor initial mass; (5) wind scaling factor; (6) ejecta mass; (7) hydrogen-rich envelope mass; (8) total mass of hydrogen; (9) progenitor radius; (10) explosion energy; (11) $^{56}$Ni mass; (12) $^{56}$Ni mixing as a fraction of the pre-SN mass.}
\end{table*}